# Multi-Branch Convolutional Network and LSTM-CNN for Heart Sound Classification

Seyed Amir Latifi[1], Hassan Ghassemian[2*], Maryam Imani[3]
[1]amir_latifi@modares.ac.ir ,[2]ghassemi@modares.ac.ir,[3]maryam.imani@modares.ac.ir
Image Processing and Information Analysis Lab, Faculty of Electrical and Computer Engineering
Tarbiat Modares University
Tehran, Iran

## Abstract

Cardiovascular diseases represent a leading cause of mortality worldwide, necessitating accurate and early diagnosis for improved patient outcomes. Current diagnostic approaches for cardiac abnormalities often present challenges in clinical settings due to their complexity, cost, or limited accessibility. This study develops two deep learning architectures that offer fast, accurate, and cost-effective methods for automatic diagnosis of cardiac diseases, focusing specifically on addressing the critical challenge of limited labeled datasets in medical contexts. We propose two methodologies: first, a Multi-Branch Deep Convolutional Neural Network (MBDCN) that emulates human auditory processing by utilizing diverse convolutional filter sizes and power spectrum input for enhanced feature extraction; second, a Long Short-Term Memory-Convolutional Neural (LSCN) model that integrates LSTM blocks with MBDCN to improve time-domain feature extraction. The synergistic integration of multiple parallel convolutional branches with LSTM units enables superior performance in heart sound analysis. Experimental validation demonstrates that LSCN achieves multiclass classification accuracy of 89.65% and binary classification accuracy of 93.93%, significantly outperforming state-of-the-art techniques and traditional feature extraction methods such as Mel Frequency Cepstral Coefficients (MFCC) and wavelet transforms. A comprehensive 5-fold cross-validation confirms robustness of our approach across varying data partitions. These findings establish the efficacy of our proposed architectures for automated heart sound analysis, offering clinically viable and computationally efficient solutions for early detection of cardiovascular diseases in diverse healthcare environments.

## 1. Introduction

Cardiovascular diseases (CVDs) remain a leading global cause of morbidity and mortality, necessitating early and accurate detection through accessible diagnostic tools to improve patient outcomes. Phonocardiography (PCG), a non-invasive and cost-effective method, offers a valuable approach to diagnose cardiac abnormalities by analyzing acoustic patterns of heart sounds, which reflect the condition of heart valves and vascular structures [1]. Despite its significant clinical potential, accurate PCG signal classification presents substantial challenges. The inherently non-stationary nature of heart sounds, coupled with variations in recording conditions and, critically, the scarcity of large, properly labeled medical datasets, collectively necessitate advanced signal processing and machine learning techniques specifically optimized for reliable use in diverse clinical settings.

While classical machine learning methods have achieved valuable results for heart sound classification, the limitations of manual feature extraction methods have led researchers to develop deep learning-based networks that demonstrate superior ability in automatic extraction of discriminative features. In the following sections, the background, motivations, and contributions of this work are presented.

### *1.1 Background and Limitations of Existing Approaches*

Current approaches to heart sound classification exhibit several limitations that constrain their widespread clinical utility. Traditional methodologies predominantly rely on manual feature extraction techniques, such as Mel Frequency Cepstral Coefficients (MFCCs), combined with statistical classifiers. While these approaches offer interpretability advantages, they typically achieve only modest accuracy levels (often 86-87% on PhysioNet/CinC datasets) and require labor-intensive feature engineering that frequently fails to capture the complex dynamics of cardiac signals [2, 3]. Conventional machine learning techniques have been explored across various cardiovascular applications; however, their effectiveness for automated heart sound classification remains fundamentally limited by the inherent challenges posed by the non-stationary nature of PCG signals.

More recently, deep learning approaches, particularly Convolutional Neural Networks (CNNs), have demonstrated superior capabilities in automated feature extraction from raw data, leading to improved classification performance (typically achieving 87-88% accuracy in recent PhysioNet/CinC challenges) [4, 5]. However, standard CNN architectures present notable limitations. They exhibit heavy dependency on large volumes of labeled data—a resource often scarce in specialized medical contexts—and their substantial computational requirements can hinder practical deployment in resource-constrained clinical environments [6]. Furthermore, most existing deep learning models for heart sound analysis focus predominantly on either time-domain or frequency-domain features in isolation, rarely integrating both modalities effectively for comprehensive signal characterization crucial for robust diagnosis.

### *1.2 Research Gap and Motivation*

The identified limitations of existing approaches highlight a significant research gap in automated heart sound classification. Traditional methods, relying on hand-crafted features, struggle to adequately model the complex and non-stationary characteristics of PCG signals, resulting in suboptimal capture of relevant diagnostic information [2, 3]. Simultaneously, while deep learning models offer enhanced feature learning capabilities, their effectiveness is often constrained by computational demands and extensive labeled dataset requirements [5].

A particularly critical gap in current literature is the lack of models that effectively integrate time-frequency analysis to simultaneously address both signal variability and the challenge of limited data availability. Such an integrated approach is essential for accurately detecting subtle cardiac abnormalities such as murmurs [3]. Moreover, the increasing global burden of cardiovascular diseases, especially in underserved regions, underscores an urgent need for reliable, cost-effective, and scalable diagnostic tools that require minimal dependence on large labeled datasets to facilitate accessible cardiac screening programs. These pressing challenges collectively provide strong motivation for developing novel deep learning architectures designed to leverage the complementary strengths of both convolutional and recurrent neural networks while mitigating their respective limitations in heart sound analysis applications.

### *1.3 Contributions*

To address the aforementioned critical challenges in heart sound-based cardiac disease diagnosis, particularly the need for accurate, cost-effective, and data-efficient methods, this study proposes two novel deep learning architectures: the Multi-Branch Deep Convolutional Neural Network (MBDCN) and the Long Short-Term Memory-Convolutional Neural (LSCN) model.

The MBDCN architecture is specifically engineered to significantly enhance frequency-domain feature extraction from heart sound signals. Drawing inspiration from the parallel processing mechanisms of the human auditory system, it employs multiple parallel convolutional branches, each utilizing diverse filter sizes, operating on Welch's periodogram inputs. This multi-branch structure allows the network to simultaneously capture spectral features across various frequency resolutions, thereby substantially improving feature discriminability and robustness. Building upon the powerful frequency-domain analysis capabilities of the MBDCN, the LSCN model integrates this architecture with Long Short-Term Memory (LSTM) units. This synergistic combination enables the network to analyze spectral content while effectively modeling crucial temporal dependencies and sequential patterns inherent within heart sound signals [7]. The LSCN thus forms a comprehensive framework capable of jointly processing and analyzing both spectral and temporal characteristics for more accurate and robust heart sound classification.

While previous studies have explored standard CNN-based models for heart sound classification, they frequently overlook or inadequately capture the intricate time-frequency relationships that are indispensable for robust signal interpretation [8]. Our proposed MBDCN and LSCN models are specifically designed to overcome this fundamental limitation through the fusion of multi-branch convolutional layers for spectral analysis and advanced temporal modeling via LSTM units. This integrated approach is expected to lead to significantly improved classification accuracy and enhanced resilience against common challenges such as signal variability and data scarcity.

Our prior investigations into traditional feature extraction methods, including Mel Frequency Cepstral Coefficients (MFCCs) and wavelet transforms, alongside entropy-based filter banks such as the Maximum Entropy Gabor Filter Bank (MEGFB) and the Maximum Entropy Mel Filter Bank (MEMFB) from our previous work [9], provided valuable insights into the critical role of effective spectral representation for heart sound analysis. MEMFB demonstrated superior performance by emulating human auditory frequency selectivity to effectively capture vital spectral regions. Building upon these foundational insights, the MBDCN and LSCN architectures were developed to leverage deep learning for automated, end-to-end feature learning and classification, addressing the fundamental challenges of data scarcity and signal variability more effectively than prior approaches.

This research contributes significantly to the field of automated heart sound analysis with the following key contributions:

- **Development of a Novel Multi-Branch CNN (MBDCN):** We introduce the MBDCN architecture, which utilizes parallel convolutional branches for robust spectral feature processing from Welch's periodogram inputs. This novel approach achieves enhanced feature extraction efficiency and demonstrates strong performance (89.15% accuracy for multiclass classification) compared to conventional CNN architectures and traditional feature-based methods.
- **Integration of LSTM for Comprehensive Time-Frequency Analysis (LSCN):** We propose the LSCN model, a novel integrated framework that synergistically combines the MBDCN's spectral analysis capabilities with LSTM units. This powerful combination effectively captures both frequency and temporal features, leading to state-of-the-art multiclass classification performance (89.65% accuracy) across variable-length heart sound recordings.
- **Demonstrated High Performance in Clinically Relevant Binary Classification for Comparative Analysis:** To facilitate a direct and fair comparison with existing state-of-the-art methods in the literature, many of which are primarily evaluated on a binary classification task, we also rigorously assessed and report the performance of our proposed LSCN model on a derived Normal vs. Abnormal classification. This approach yielded exceptional performance metrics: 93.93% accuracy, 94.77% sensitivity, and 92.96% specificity, highlighting LSCN's strong potential as a reliable screening tool and demonstrating its competitive edge against models optimized solely for binary classification.
- **Robust Validation on Benchmark Datasets:** The proposed models are rigorously validated through comprehensive 5-fold cross-validation conducted on challenging, imbalanced benchmark datasets (PhysioNet/CinC 2016 and 2017 [10, 11]), confirming their robustness, generalizability, and reliability across diverse data partitions and real-world conditions.
- **Potential for Deployment in Low-Resource Settings:** The comprehensive experimental evaluation demonstrates the high effectiveness of our proposed models, particularly the LSCN in the critical binary classification scenario, on established benchmark datasets. This establishes their promise as computationally viable and clinically applicable solutions for deployment in low-resource healthcare environments.

## *1.4 Related Works*

The classification of heart sounds for cardiovascular disease diagnosis has evolved through distinct methodological phases, each addressing fundamental challenges in automated cardiac assessment.

### *1.4.1 Classical Feature-Based Approaches (2010-2018)*

Early research in heart sound classification relied heavily on handcrafted feature extraction combined with traditional machine learning algorithms. These methods typically employed spectral and time-domain features such as Mel-Frequency Cepstral Coefficients (MFCCs) [12], Discrete Wavelet Transforms (DWT), filter bank representations including Maximum Entropy Gabor Filter Bank (MEGFB) and Maximum Entropy Mel Filter Bank (MEMFB) [9]. These handcrafted features were often combined with classifiers such as Support Vector Machines (SVM), Decision Trees, or K-Nearest Neighbors (KNN). While computationally efficient and interpretable, these approaches were constrained by their dependency on manual feature engineering and limited generalizability across diverse datasets, typically achieving accuracies in the range of 86-87% on standard benchmarks like PhysioNet/CinC.

### *1.4.2 Deep Learning Transition (2018-2022)*

The advent of deep learning marked a significant paradigm shift toward end-to-end feature learning in heart sound analysis. Ren et al. [13] pioneered application of deep CNNs by converting phonocardiogram (PCG) signals into scalogram-based image representations, employing VGG architecture with SVM classification. However, their approach achieved limited performance (56.2% accuracy), highlighting the challenges of adapting image-based techniques to acoustic signal processing and indicating the complexities involved in converting temporal signals to image representations.

Building upon this foundation, Li et al. [14] proposed a hybrid methodology that combined handcrafted features with convolutional neural networks, achieving improved performance (86.8% accuracy) while maintaining some reliance on manual feature design. This work demonstrated the potential benefits of integrating traditional feature extraction with deep learning architectures, though the dependence on manual feature design limited adaptability.

Zhou et al. [15] introduced Dense-FSNet, a novel 1D-CNN-based architecture incorporating densely connected feature selection mechanisms. Their model achieved competitive performance (86.09% accuracy, 88.5% sensitivity) by leveraging direct connections between layers to enhance feature propagation and reduce information loss. This approach represented a significant advancement in purely deep learning-based heart sound classification, demonstrating the effectiveness of 1D-CNN architectures for temporal signal processing.

More recently, Riccio et al. [16] explored the integration of fractal analysis with CNN architectures for phonocardiogram classification. Despite the innovative approach combining geometric complexity measures with deep learning, their method achieved moderate performance (70% accuracy), indicating the ongoing challenges in effectively capturing the complex characteristics of cardiac acoustic signals and highlighting the limitations of single-domain feature extraction approaches.

#### *1.4.3 Temporal-Spectral Integration in Biomedical Signals (2020-Present)*

Contemporary research has increasingly recognized the importance of modeling both spatial (frequency-domain) and temporal dynamics in biomedical signal analysis. Hybrid architectures combining Convolutional Neural Networks with Long Short-Term Memory (LSTM) networks have demonstrated substantial promise across various biomedical applications, establishing the foundation for our proposed approach.

Srivastava et al. [17] introduced ApneaNet, a hybrid 1D-CNN-LSTM architecture for obstructive sleep apnea detection from digitized ECG signals. Their model effectively combined convolutional feature extraction with sequential modeling via LSTM layers, achieving robust performance even on imbalanced datasets. The architecture demonstrated the effectiveness of temporal-spectral fusion for biomedical signal classification, where both spatial (frequency-domain) and temporal dynamics are essential for accurate diagnosis.

Similarly, Efe and Ozsen [18] proposed CoSleepNet for automated sleep stage classification using EEG and EOG signals, leveraging CNNs for local pattern detection and LSTMs for capturing temporal dependencies across sleep cycles. Their hybrid CNN-LSTM network successfully handled imbalanced datasets and demonstrated superior performance compared to individual CNN or LSTM architectures, validating the effectiveness of dual-domain approaches for complex biomedical signal analysis.

These studies demonstrate the effectiveness of temporal-spectral fusion architectures in biomedical signal classification, where the combination of convolutional layers for feature extraction and LSTM networks for temporal modeling provides superior performance compared to single-domain approaches.

#### *1.4.4 Contemporary AI-Driven Medical Diagnostics*

Recent advances highlight significant progress in AI-driven medical diagnostics, particularly for cardiac and respiratory conditions, demonstrating the broader applicability of hybrid deep learning approaches in healthcare. Tartarisco et al. [19] developed Machine-Cyber-Physical Systems (M-CPS) to facilitate heart valve disorder screening in resource-limited settings, showcasing the potential for AI-enabled diagnostic systems in challenging clinical environments where traditional diagnostic equipment may be unavailable or cost-prohibitive.

Zhang et al. [20] demonstrated the effectiveness of dual-channel CNN-LSTM algorithms in classifying lung sounds for pulmonary assessment, further validating the CNN-LSTM hybrid approach for respiratory signal analysis. Their work reinforced the importance of combining spatial feature extraction capabilities of CNNs with temporal modeling strengths of LSTMs for accurate biomedical signal classification, particularly in scenarios involving complex acoustic patterns.

Zhou et al. [21] advanced pediatric cardiac diagnostics by combining vision transformers with Gramian angular field representations to distinguish between normal, innocent, and pathological pediatric heart sounds. Their innovative approach demonstrated the potential of transformer-based architectures in cardiac acoustic analysis, while highlighting the ongoing need for specialized preprocessing techniques to effectively capture the temporal-spectral characteristics of heart sounds.

These studies illustrate the accelerating integration of AI techniques in healthcare diagnostics and inform our approach, which builds upon similar hybrid architectural principles (e.g., CNN-LSTM integration in LSCN) while maintaining a specific focus on heart sound analysis with emphasis on data efficiency and computational accessibility for real-world clinical deployment.

#### *1.4.5 Summary of Literature Limitations*

Despite these advances, existing methods continue to face several critical limitations: (1) dependency on large labeled datasets, which are often scarce in medical domains; (2) sensitivity to signal variability and recording conditions; (3) computational complexity limiting deployment in resource-constrained clinical environments; and (4) insufficient modeling of both spectral and temporal characteristics inherent in cardiac acoustic signals, particularly in low-resource settings.

These limitations, combined with the insights gained from our comprehensive literature review, reinforce the motivation for developing the proposed architectures described in Section 1.2 and detailed in Section 1.

### *1.5 Research Objectives*

The primary objectives driving this research are:

- To design and develop the novel MBDCN architecture for advanced frequency-domain feature extraction from heart sounds.
- To propose and implement the LSCN model by integrating the MBDCN with LSTM units to enhance time-frequency analysis capabilities.
- To conduct a thorough experimental evaluation and comparative analysis of the proposed MBDCN and LSCN models against established state-of-the-art methods using the challenging PhysioNet/CinC 2016 and 2017 datasets [10, 11], aiming to demonstrate superior and robust classification performance for both multiclass and derived binary tasks.

Despite these significant advancements, certain challenges persist. The computational complexity of the LSCN model, while yielding high accuracy, may present limitations for immediate deployment in severely resource-constrained settings. Furthermore, the inherent "black-box" nature of deep learning models can reduce interpretability compared to traditional methods, potentially posing barriers to clinical acceptance. Future research will focus on optimizing computational efficiency (e.g., via model pruning or quantization) and developing techniques to improve the interpretability of these models to facilitate their practical clinical adoption.

The remainder of this paper is organized as follows: Section 2 describes the proposed and baseline methods. Section 3 presents experimental results. Section 4 provides discussion of findings, and Section 5 concludes the study.

## 2. METHODS

This section presents two novel deep learning architectures designed for heart sound classification. We also describe the baseline methods used for comparison. Comprehensive details are provided regarding the dataset, preprocessing steps, evaluation metrics, and training procedures to ensure reproducibility and clarity.

### *2.1 PROPOSED METHODS*

We introduce two innovative architectures for heart sound classification: 1) the Multi-Branch Deep Convolutional Neural Network (MBDCN) and 2) the Long Short-Term Memory-Convolutional Neural (LSCN) model. These methods combine human auditory-inspired feature extraction with deep learning techniques to address challenges in cardiac signal analysis, particularly data scarcity and signal variability. We present their designs, theoretical foundations, and implementation algorithms in detail.

#### *2.1.1 The Multi Branch Deep Convolutional Neural (MBDCN) network*

**Architecture Overview and Theoretical Foundation:**

The MBDCN architecture implements a biomimetic multi-scale convolutional framework that emulates frequency-selective mechanisms in the human cochlea [22]. This approach is motivated by the cochlea's ability to process multiple frequency components in parallel, enabling efficient spectral decomposition of complex acoustic signals. Our implementation utilizes a one-dimensional Deep Convolutional Neural Network (DCNN) with multiple parallel branches, each extracting different levels of information from the input data using specifically calibrated filter sizes. For audio signal analysis, particularly heart sounds, this parallel branch approach offers significant advantages over traditional sequential architectures. The human auditory system employs a wave bank filter mechanism to capture specific frequency components in audio signals, with different regions of the basilar membrane responding to distinct frequency ranges. By incorporating parallel branches with carefully selected filter sizes (ranging from 1×3 to 1×37), the MBDCN network similarly detects patterns within specific frequency ranges, enabling multiscale feature representation critical for capturing different aspects of heart sounds. Unlike conventional multi-branch architectures developed for image processing or general signal analysis, the MBDCN specifically combines Welch's periodogram inputs with a tailored set of filter sizes optimized for cardiac acoustic signatures [23]. This unique combination enhances feature extraction specificity for heart sound signals, improving discrimination between normal and pathological patterns. The general block diagram of the MBDCN network is illustrated in Figure 1.

The MBDCN architecture is designed to capture cardiac sound patterns across multiple temporal and spectral resolutions through a parallel multi-branch convolutional framework. Each branch is assigned a specific receptive field scale, enabling the network to extract complementary features ranging from fine-grained details to global structural patterns.

**Multi-Resolution Rationale:** Inspired by wavelet decomposition principles, the architecture processes the same spectral input in parallel convolutional streams with different kernel sizes. Small kernels focus on localized, high-frequency variations, while large kernels capture broader, low-frequency trends. This combination enables multi-scale feature extraction within an end-to-end trainable deep learning framework.

**Biological Analogy:** The human auditory system, specifically the cochlea, employs multiple frequency-selective filters operating in parallel, each tuned to a different frequency range. MBDCN mimics this mechanism by employing convolutional kernels with varied receptive fields to analyze the frequency-domain representation of heart sounds obtained through spectral analysis.

Branch-wise Kernel Design: The network comprises four parallel convolutional branches:

- Branch 1 (Small-scale focus): 1×3 → 1×3 → 1×3, designed to capture rapid spectral transitions and high-frequency components.
- Branch 2 (Medium-scale focus): 1×11 → 1×9 → 1×7, extracting intermediate spectral features.
- Branch 3 (Large-scale focus): 1×21 → 1×15 → 1×11, modeling extended frequency structures.
- Branch 4 (Very large-scale focus): 1×37 → 1×25 → 1×17, capturing long-range dependencies and global spectral envelopes.

**Empirical Optimization:** Kernel sizes were determined through systematic ablation studies, where different kernel size combinations were tested while keeping all other architectural parameters constant to evaluate their isolated impact on performance. Through this trial-and-error approach, we identified that the configuration spanning from fine-grained (1×3) to global (1×37) analysis scales provides optimal balance between comprehensive spectral coverage and computational efficiency, ensuring adequate representation of both narrow-band and wide-band spectral characteristics essential for cardiac acoustic pattern recognition.

**Hierarchical Feature Extraction:** Within each branch, three successive convolutional layers progressively increase the number of filters (22 → 32 → 64) while reducing kernel size to refine feature granularity. Batch normalization and ReLU activation follow each convolution to enhance convergence stability and non-linearity. The outputs from all branches are concatenated to form a rich, multi-scale feature representation for downstream classification.

**MBDCN Workflow and Implementation:**

The workflow of the MBDCN is formalized in Algorithm 1, detailing the step-by-step process from input preprocessing to classification.

**Algorithm 1: MBDCN Workflow**

1. **Input Preprocessing**
   - Compute the power spectrum using Welch's method [24]: segment the signal into overlapping 5-second intervals, apply a Hamming window, and average the power spectral density (PSD) to reduce noise.
   - Reshape the PSD into a four-dimensional (4D) tensor (1 × number of features × 1 × number of samples) for network input.
2. **Multi-Branch Convolution**
   - Develop four parallel convolutional branches utilizing filter sizes ranging from 1×3 to 1×37 across distinct convolutional pathways. Employ kernel counts of 32 and 64 to extract both local and global features effectively.
   - Apply batch normalization and ReLU activation to each branch output, ensuring training stability and non-linearity.
3. **Feature Integration**
   - Concatenate outputs from all branches using a depth concatenation (Depthcat) layer to construct a unified multi-scale feature vector.
   - Refine the feature vector using a 1D convolutional layer (filter size 1×3) and max pooling to reduce dimensionality while preserving essential patterns.
4. **Classification**
   - Process the final feature vector through a fully connected layer and softmax activation to classify signals into normal or abnormal categories.

The mathematical formulation for the multi-branch feature extraction can be expressed as:

$$F_{mb} = Concat\left(f_1(X), f_2(X), f_3(X), f_4(X)\right) \tag{1}$$

where $F_{mb}$ represents the concatenated multi-branch features, $f_i(X)$ denotes the feature maps from the $i$-th branch with its specific filter size, and $X$ is the input power spectrum. Each branch function $f_i$ encompasses convolution, batch normalization, and ReLU activation:

$$f_i(X) = ReLU\left(BN\left(Conv_{ki}(X)\right)\right) \tag{2}$$

where $ki$ represents the kernel size for branch $i$.

The MBDCN's multi-branch architecture, illustrated in Figure 1, automates feature extraction across multiple scales, minimizing dependence on manual feature engineering while effectively capturing hierarchical patterns in heart sound signals. This biologically inspired framework enhances the efficiency of deep learning models in biomedical signal analysis, offering a robust solution for automated cardiac diagnosis.

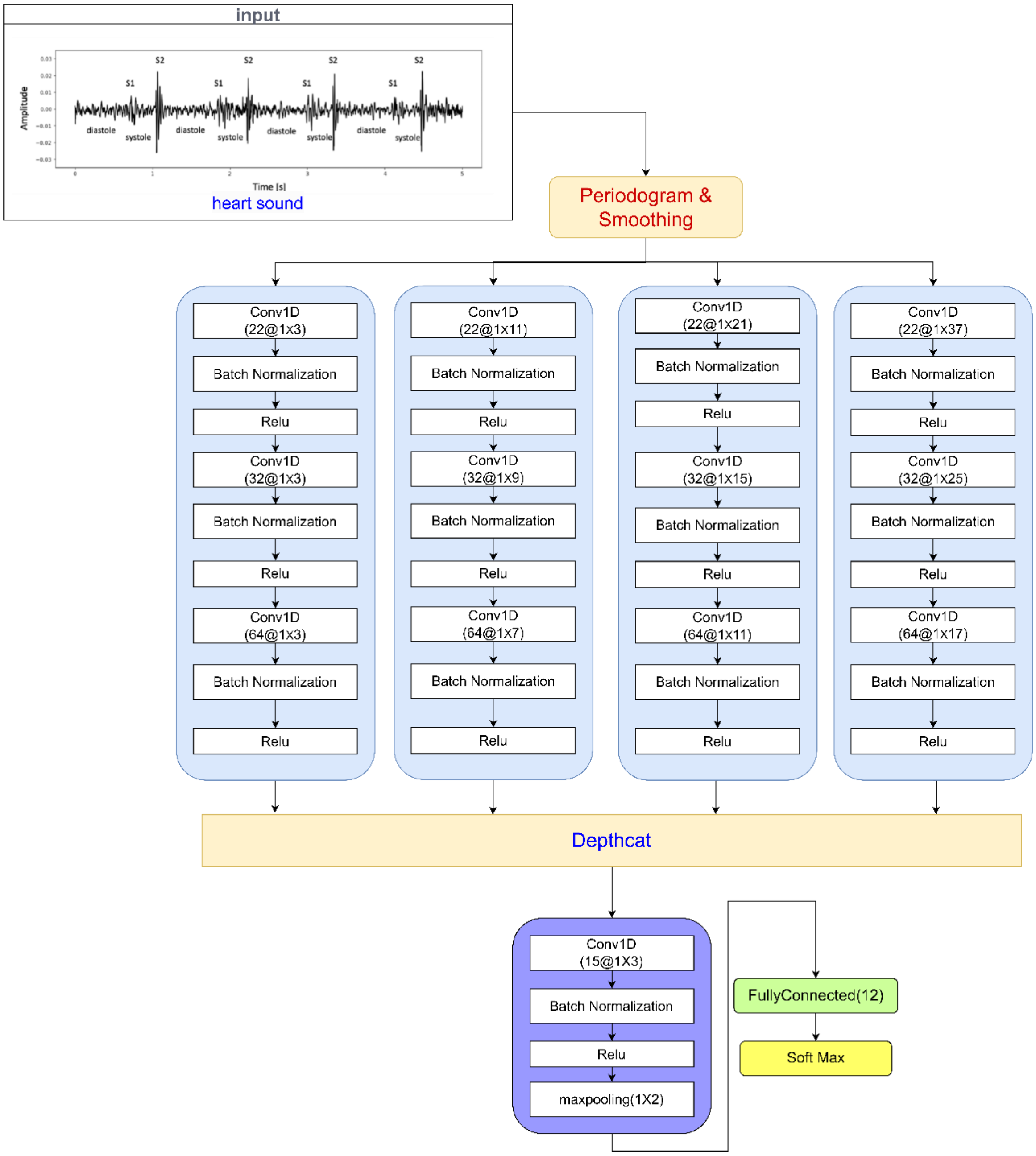

**Figure 1.** Architecture of the proposed Multi-Branch Deep Convolutional Network (MBDCN) for heart-sound classification. Input signals undergo periodogram analysis and smoothing to enhance spectral visibility. Multiple parallel 1D convolutional branches with varying kernel sizes (1×3 to 1×37) extract multi-scale spectral–temporal features; each branch includes batch normalization and ReLU activation. Branch outputs are depth-concatenated and further processed by additional convolutional layers, max-pooling, and a fully connected layer with softmax for classification. The architecture features parallel convolutional branches to capture multi-scale features, with clear labels indicating the spectral processing pathway

### *2.1.2 . Long Short-term memory-Convolutional Neural (LSCN) network:*

**Architecture Overview and Theoretical Basis:**

The proposed LSCN model extends the MBDCN architecture by integrating Convolutional Neural Network (CNN) layers for frequency-domain feature extraction with Long Short-Term Memory (LSTM) components for sequential learning in the time domain [25]. This architecture terminates in a softmax classifier for binary classification of heart sounds. The LSCN represents a significant advancement over typical hybrid models by synergistically combining spectral features extracted through MBDCN's Welch's method-based multi-branch convolutions with temporal features captured via strategically configured LSTM blocks. This dual-domain approach facilitates comprehensive learning of complex heart sound patterns across both frequency and time dimensions. The modified MBDCN network architecture incorporated within the LSCN framework is illustrated in Figure 2.

LSTM networks [17] were specifically developed to address limitations in traditional recurrent neural networks (RNNs), particularly their inability to effectively model long-term dependencies in sequential data. The LSTM architecture consists of three primary structural components: input, hidden, and output layers. The hidden layer contains memory blocks with memory cells regulated by three specialized gating mechanisms: the forget gate, input gate, and output gate. These gates collaboratively govern information flow through the network, selectively preserving relevant temporal information while discarding irrelevant data. This capability is particularly valuable for heart sound analysis, where temporal patterns often span multiple cardiac cycles.

CNNs excel at hierarchical feature extraction through convolutional operations but inherently lack the ability to model sequential dependencies effectively. They employ stacked convolutional blocks with non-linear activation functions to extract spatial patterns, followed by pooling operations to distill salient features while reducing dimensionality. However, their fundamental limitation lies in capturing long-range temporal dependencies, a critical requirement for accurate heart sound classification.

The LSCN framework capitalizes on the complementary strengths of both architectures: CNN's capacity for robust local feature extraction and LSTM's proficiency in sequential pattern recognition. By incorporating dual LSTM units (600 and 100 units respectively), the model efficiently processes time-series cardiac data while leveraging CNN's ability to detect intricate frequency-domain features. This architectural integration significantly enhances performance in tasks requiring comprehensive analysis of biomedical signals with complex temporal-spectral characteristics [26]. The LSCN features a dual-pathway design: raw temporal data feeds directly into the LSTM stream (preserving sequence information), while spectral features are extracted via the MBDCN stream from Welch's periodogram inputs.

**LSCN Workflow and Implementation:**

**Algorithm 2: LSCN Workflow**

**1. Input Preprocessing**

- Apply Welch's method to convert time-domain signal into frequency domain for spectral analysis.
- Two parallel preprocessing streams are established: raw temporal data for LSTM (processed directly as sequences to preserve temporal structure) and spectral data for CNN.
- Tensor dimensions are as follows: for the raw temporal data, the shape is defined as (1 × time_steps × 1 × number of samples) which is then reshaped (num_samples, time_steps, 1) for LSTM networks; whereas for the spectral data, after applying the Welch method, the shape is represented as (1 × frequency_bins × 1 × number of samples).

**2. Dual-Stream Feature Extraction**

- **Spectral Stream:** Extract frequency-based features using MBDCN's multi-branch CNN layers from Welch periodogram.
- **Extract temporal dependencies using two LSTM blocks Temporal Stream**:
    - First LSTM block: 600 units
    - Second LSTM block: 100 units

**3. Feature Integration**

- **CNN Features:** Multi-branch CNN outputs are flattened to prepare for integration.

- **LSTM Features:** LSTM outputs from temporal processing are prepared for integration.
- **Concatenation Layer:** Flattened CNN spectral features and LSTM temporal features are combined using a concatenation layer to create a unified multi-modal feature vector.
- The concatenated features are processed through a fully connected layer.

**4. Classification**

- Utilize a softmax classifier for final binary classification (normal vs. abnormal).

This workflow is illustrated in Figure 2, showing the parallel streams and concatenation.

The LSCN framework effectively integrates CNNs for spatial feature extraction and LSTMs for sequential learning through a sophisticated dual-stream architecture. The key innovation lies in the concatenation-based fusion of CNN spectral features and LSTM temporal features, enabling robust classification of biomedical signals. This architecture capitalizes on the complementary strengths of both models, offering superior performance in analyzing complex time-series data by processing both frequency-domain and time-domain characteristics simultaneously.

The mathematical representation of the LSCN model can be formulated as:

$$F_{CNN} = MBDCN(X) \quad (3)$$

$$F_{LSTM} = LSTM_{100}\left(LSTM_{100}(F_{CNN})\right) \quad (4)$$

$$\hat{y} = softmax(W \cdot F_{LSTM} + b) \quad (5)$$

where $F_{CNN}$ represents features extracted by the MBDCN component, $F_{LSTM}$ denotes the temporal features after LSTM processing, and $\hat{y}$ is the final classification output. $W$ and $b$ are the weight matrix and bias vector of the fully connected layer, respectively.

The specific configuration of LSTM units (600 followed by 100) was determined through extensive experimentation and hyperparameter optimization. The first layer with 600 units captures detailed temporal relationships in the data, while the second layer with 100 units distills this information into a more compact representation, preventing overfitting while preserving discriminative temporal features. This architecture effectively integrates CNNs for spatial feature extraction and LSTMs for sequential learning, enabling robust classification of complex biomedical signals.

.

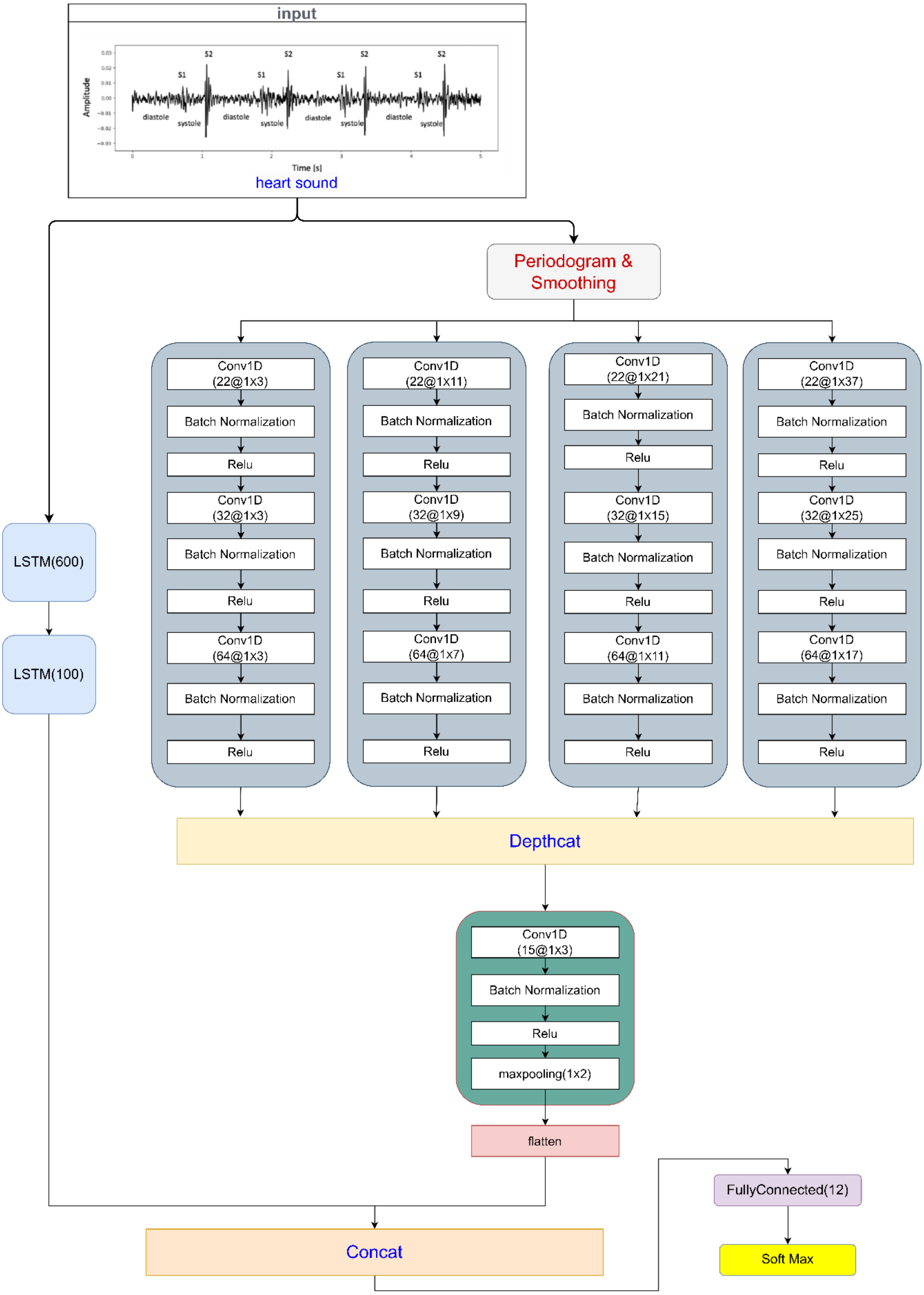

**Figure 2**. Architecture of the proposed LSCN for heart sound classification. The model employs dual-stream processing: (1) spectral analysis through MBDCN with multiple 1D-CNN branches, and (2) temporal modeling using LSTM layers (600 and 100 units). Features from both streams are concatenated and passed through fully connected layers for final classification

#### 2.1.3 Implementation Details:

Both models were developed and trained on the PhysioNet/CinC 2016 dataset [11]. The design process for the 1D-CNN components required iterative optimization of various configurations and meta-parameters, including learning rate, batch size, and filter dimensionality, to determine the optimal architecture for heart sound classification. This process involved training multiple CNN models with different configurations and conducting systematic meta-parameter searches to identify the most effective filter sizes and architectural organization for cardiac acoustic analysis.

Developing an effective one-dimensional CNN for cardiac audio processing necessitated careful consideration of network architecture, training methodology, and data preprocessing techniques. Through extensive experimentation, we optimized both models to accurately capture the distinctive spectral and temporal features of heart sound signals, enabling effective discrimination between normal and pathological patterns.

To provide comprehensive validation of our proposed methodologies and demonstrate their superior performance compared to traditional machine learning approaches, we implement and evaluate these methods in the subsequent experimental section.

### 2.2 Competing methods

To evaluate our proposed MBDCN and LSCN architectures, we implement several established baseline methods. These competing approaches include both traditional machine learning and contemporary deep learning techniques commonly used in heart sound classification.

#### 2.2.1 .1D-CNN Network

The One-dimensional CNN models are capable of extracting features from input data using different filter sizes. These networks consist of convolutional blocks designed to extract hierarchical patterns, followed by fusion layers that refine and integrate extracted features. By leveraging shared information across layers, a 1D-CNN creates a feature-rich representation at a lower computational cost compared to multidimensional transformations. Fully connected layers, coupled with nonlinear activation functions such as SoftMax or Sigmoid, enable effective classification. Additionally, batch normalization and dropout layers are incorporated to prevent overfitting [18].

During training, network kernels are either initialized randomly or using transfer learning techniques. Optimization algorithms such as Adaptive Moment Estimation (Adam) or Stochastic Gradient Descent with Momentum (SGDM) are employed to fine-tune the model. One key advantage of 1D-CNNs is their ability to efficiently extract local features while maintaining computational efficiency, making them particularly suitable for heart sound classification. For heart sound classification, we designed a 1D-CNN model, illustrated in Figure 3. The architecture consists of four convolutional layers with a kernel size of 3×1, followed by ReLU activation and max pooling (2×1). A fully connected (FC) layer with 12 neurons and a softmax output function classifies heart sounds into 12 distinct categories, including normal and abnormal patterns. The Welch periodogram was used for spectral analysis, and its power spectrum was fed into the 1D-CNN network.

**Architecture Overview:**

- **Input:** Power spectrum derived from Welch's method.
- **Layers:** Four 1D convolutional layers (kernel size: 3×1, channels: 22, 32, 64, 86), each with ReLU activation, followed by max pooling (2×1) and a softmax output layer for 12-class classification.
- **Training:** Optimization using SGDM, with batch normalization to prevent overfitting.
- **Strengths and Limitations:** While 1D-CNN efficiently extracts local features with moderate computational cost, it lacks the temporal modeling capabilities provided by LSTM components, a limitation addressed in our LSCN framework.

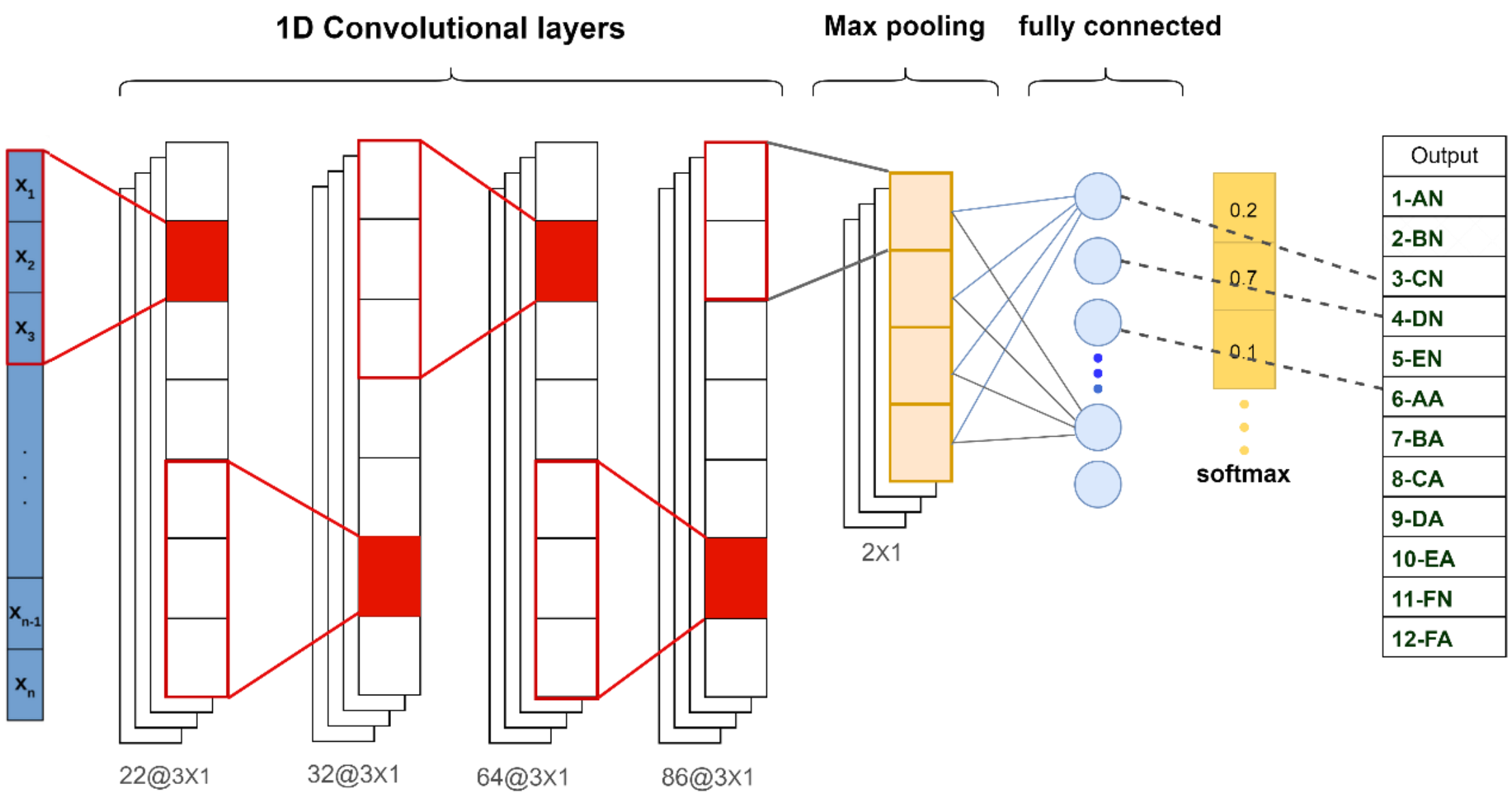

Figure 3. 1D-CNN architecture. The block diagram shows that from the input, successive convolution layers are followed by a max pooling layer with a window of size (2 × 1). The network has four convolution layers with a kernel size of (3 × 1), while the channel depth goes successively as 22, 32, 64, and 86. These are further followed by a fully connected layer terminated with a softmax layer that outputs the probabilities over the 12 possible classes ranging from 1-AN to 12-FA.

### *2.2.2 . Multilayer Perceptron (MLP) Baseline*

The MLP architecture provides a fully connected neural network baseline for performance comparison, representing the simplest deep learning approach to heart sound classification.

**Architecture Overview:**

- **Input:** Preprocessed features (e.g., MFCCs or raw signals).
- **Network Structure:** Six fully connected layers with ReLU activation
- **Output**: Softmax classification layer for multi-class prediction
- **Training:** Stochastic Gradient Descent with Momentum (SGDM) and dropout regularization
- **Comparative Analysis:** Despite its architectural simplicity facilitating implementation and interpretation, the MLP fundamentally lacks the capacity to model spatial or temporal patterns effectively, unlike the specialized architectures of MBDCN and LSCN.

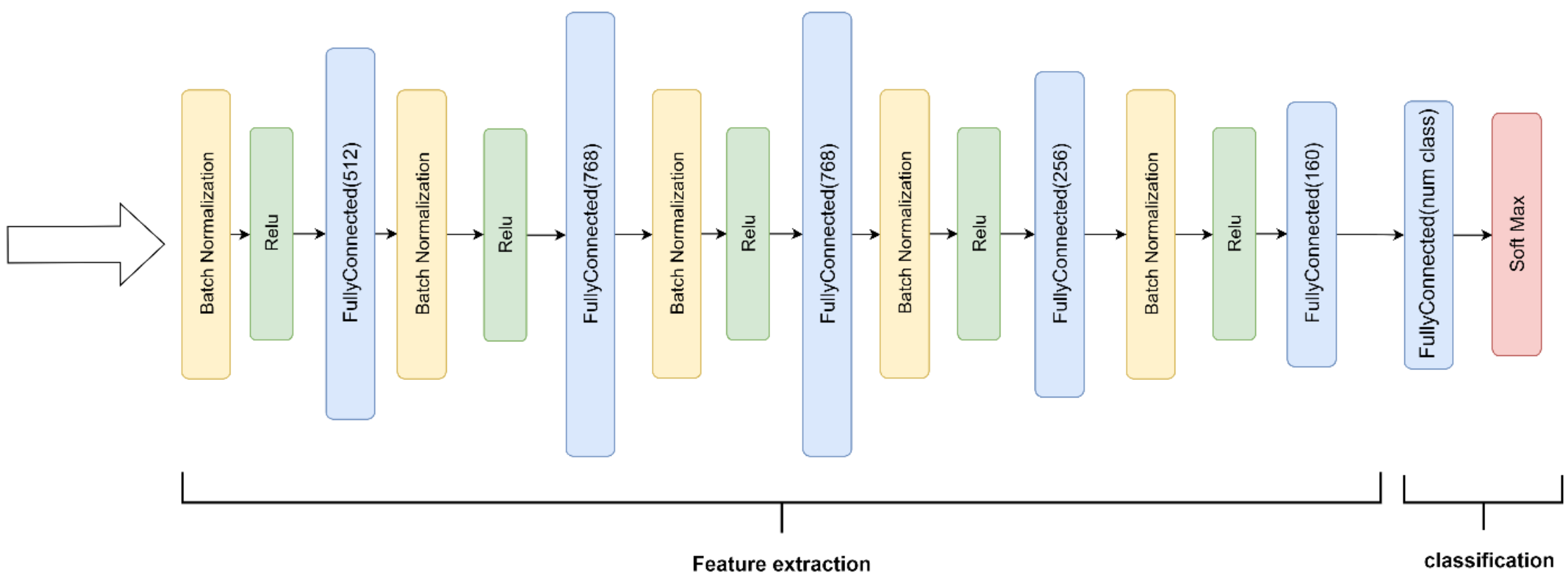

Figure 4. Block diagram of MLP network

### *2.2.3 . Classical Machine Learning Methods*

Classical machine learning approaches for heart sound classification typically combine manual feature extraction with traditional classifiers, offering interpretable solutions for biomedical signal analysis [9, 27].

This section evaluates these methods, focusing on their feature extraction mechanisms, classifier performance, and application to cardiac acoustic data.

**Feature Extraction and Classification Pipeline:**

Our analysis encompassed multiple feature extraction techniques referenced in previous research [9], including Mel-frequency cepstral coefficients (MFCC) [12], discrete wavelet transforms (DWT), Maximum Entropy Gabor Filter Bank (MEGFB), and a hybrid approach combining MEGFB with MFCC (MEMFB). These feature extraction methods were systematically evaluated in conjunction with established classification algorithms, including Maximum Likelihood (ML), Support Vector Machines (SVM) with various kernel functions, Decision Trees with adaptive pruning, and K-Nearest Neighbors (KNN) with optimized distance metrics.

For example, the combination of MEMFB features with an SVM classifier using a radial basis function kernel achieved competitive results for binary classification of heart sounds (normal vs. abnormal), though this approach required significant manual parameter tuning. Additionally, we evaluated filter bank approaches such as MFCC, WaveFB, MEMFB, and MEGFB for feature extraction from lung sound data, adapting these methodologies to provide contextual comparison for our proposed cardiac sound classification techniques. A notable implementation difference between lung and heart sound analysis emerged in the application of MEMFB and MEGFB: for the PhysioNet cardiac dataset, we utilized 26 Gabor filters to optimally capture the spectral characteristics of heart sounds.

**Comparison with Proposed MBDCN:**

To establish meaningful benchmarks for classical methods, we employed our proposed MBDCN as a classifier, contrasting its performance with traditional classification algorithms (ML, SVM, Decision Trees, and KNN). Unlike classical approaches that depend on labor-intensive feature engineering, MBDCN automates the feature extraction process through its multi-branch convolutional architecture. This automation substantially reduces the need for manual parameter tuning while enhancing adaptability to the non-stationary characteristics of heart sound signals, representing a significant advancement over traditional methodologies.

**Strengths and Limitations:**

Classical machine learning methods offer computational efficiency and interpretability advantages, making them particularly suitable for resource-constrained clinical environments. However, they face significant limitations in heart sound classification:

- **Labor-Intensive Feature Engineering:** The reliance on manual feature extraction (e.g., MFCC, DWT) necessitates extensive domain expertise and parameter tuning, limiting scalability and standardization.
- **Limited Generalization:** Classical methods often struggle to generalize effectively across diverse datasets due to their sensitivity to signal variability and recording noise, common challenges in PCG data acquisition.
- **Lower Accuracy:** As demonstrated in our experimental results (Section 4), classical approaches generally underperform compared to deep learning architectures like MBDCN and LSCN, particularly for complex classification tasks.

These limitations underscore the need for automated, robust classification solutions that can effectively address the complexities inherent in cardiac sound analysis.

### *2.2.4 . State-of-the-Art Deep Learning Methods*

For comprehensive evaluation, we implement and compare against recent state-of-the-art approaches reported in the literature:

**Ren et al. (2018) Implementation [13]:**

- **Architecture:** VGG-based CNN with scalogram image conversion
- **Input Processing:** PCG signals converted to scalogram representations
- **Classification:** SVM applied to extracted deep features
- **Training:** Transfer learning from ImageNet pretrained weights

**Li et al. (2020) Implementation [14]:**

- **Architecture:** Hybrid CNN combining handcrafted and learned features
- **Input Processing:** Manual feature extraction followed by CNN processing
- **Training:** End-to-end training with feature fusion layer
- **Optimization:** Standard backpropagation with Adam optimizer

**Zhou et al. (2021) Implementation [15]:**

- **Architecture:** Dense-FSNet with 1D-CNN and densely connected layers
- **Input Processing:** Raw heart sound signals with automatic feature selection
- **Network Design:** Densely connected feature selection mechanisms
- **Training:** Direct end-to-end learning from raw audio signals

**Riccio et al. (2023) Implementation [16]:**

- **Architecture:** CNN with fractal feature integration
- **Input Processing:** Fractal analysis combined with standard CNN features
- **Feature Extraction:** Geometric complexity measures integrated with deep learning
- **Training:** Hybrid approach combining mathematical and learned features

All methods were implemented using identical experimental conditions including dataset preprocessing, evaluation protocols, and hardware specifications to ensure fair comparative assessment.

#### *2.2.5 . Relevance to Proposed Methods*

The comparative analysis of existing approaches reveals important distinctions that highlight the contributions of our proposed methods. While 1D-CNN and MLP architectures emphasize spatial or static feature extraction, and classical methods offer simplicity and interpretability, they fundamentally lack the multi-scale analysis capabilities and temporal modeling strengths of MBDCN and LSCN. The broader landscape of AI advancements reinforces the value of hybrid model architectures, yet our specific focus on auditory-inspired multi-branch convolutions combined with LSTM integration for cardiac sound analysis represents a distinctive contribution to the field. The experimental results presented in Section 4 quantify these comparative advantages through systematic evaluation.

### *2.3.Data and Preprocessing*

Heart sound recordings were obtained from diverse participants worldwide in both clinical and non-clinical environments, including healthy individuals and patients with diagnosed cardiac conditions such as heart valve disease and coronary artery disease. The primary dataset utilized for model development and evaluation was the PhysioNet/CinC 2016 Challenge dataset [11], comprising 4,430 recordings from 1,072 subjects, collectively containing 233,512 individual heart sounds. The dataset contains two principal classes: normal and abnormal heart sounds.

The dataset originates from six distinct sources (labeled A through F), with recording durations ranging from 5 to over 20 seconds. The distribution of samples across these sources varies considerably: Dataset A contains 292 normal and 117 pathological samples; dataset B comprises 104 normal and 386 abnormal samples; dataset C includes 24 normal and 7 abnormal cases; dataset D contains 28 abnormal and 27 normal samples; dataset E consists of 183 normal and 1,958 abnormal heart sound recordings; and dataset F includes 34 normal and 80 abnormal samples. All signals were originally recorded at a sampling rate of 44.1 kHz, with durations varying between 1 and 120 seconds.

Following established practice in biomedical signal processing, we consolidated these six sources into a unified dataset to enable comprehensive evaluation across both normal and abnormal cardiac acoustic patterns. To ensure uniformity in signal analysis, we standardized each recording to a 5-second duration. Signal preprocessing included application of a Butterworth low-pass filter with a cutoff frequency of 900 Hz to eliminate DC offset and high-frequency artifacts, followed by down sampling from 44.1 kHz to 2 kHz to reduce computational requirements while preserving diagnostically relevant frequency components [28].

Energy normalization was applied to all heart sound recordings using the following formula:

$$x_{norm}[n] = \frac{x[n] - \mu_x}{\sigma_x} \; ; n = 1, \ldots, N \qquad (5)$$

where $x[n]; n = 1, \ldots, N$ is the heart's sound samples, $\mu_x$ and $\sigma_x$ are the mean and standard deviation of $x[n]$, Respectively.

It is important to note that the recordings in this dataset contain various types of environmental and physiological noise, including conversational artifacts, stethoscope movement sounds, breathing, and intestinal sounds. The dataset's diversity extends to demographic representation, encompassing both pediatric and adult populations of all genders. To address the challenge of limited training samples and ensure uniform input dimensions, each recording was segmented using overlapping 5-second windows (with a Hamming window for smoothing to enhance frequency visibility, capture continuous cardiac patterns, and reduce boundary artifacts). This segmentation process substantially expanded the available training data from 4,430 original recordings to 13,015 segments. Residual segments shorter than 5 seconds at the end of recordings were systematically discarded to maintain input consistency and avoid incomplete cardiac cycles that could introduce classification bias. Recordings ≤5 seconds were retained as single segments without further division, explaining unchanged counts in some groups (e.g., Group B). Residual segments shorter than 5 seconds were discarded to maintain data quality.

Table 1. Number of Data Entries for Each Class of Heart Sound Signals

| **Database** | **Group A** | | **Group B** | | **Group C** | | **Group D** | | **Group E** | | **Group F** | |
|---|---|---|---|---|---|---|---|---|---|---|---|---|
| **Segmentation** | AA | AN | BA | BN | CA | CN | DA | DN | EA | EN | FA | FN |
| **Original Data** | 117 | 292 | 386 | 104 | 7 | 24 | 27 | 28 | 1958 | 183 | 80 | 34 |
| **5-Second Segmentation** | 738 | 1852 | 386 | 104 | 51 | 240 | 51 | 87 | 8129 | 665 | 502 | 210 |

Table 1 provides a comprehensive overview of sample distribution across the six dataset sources (A through F), each subdivided into Normal (denoted by N) and Abnormal (denoted by A) categories. The table presents both the original data counts and the expanded dataset after 5-second segmentation, illustrating the significant increase in available training samples.

## *2.4.Evaluation Metric*

To comprehensively assess model performance, we employed multiple complementary evaluation metrics, including accuracy (AC), sensitivity (SE), specificity (SP), and precision (PR). These metrics are defined as follows:

$$Accuracy(\%) = \frac{TP + TN}{TP + TN + FP + FN} \times 100 \qquad (6)$$

$$Sensitivity(\%) = \frac{TP}{TP + FN} \times 100 \qquad (7)$$

$$Specificity(\%) = \frac{TN}{TN + FP} \times 100 \qquad (8)$$

$$Precision(\%) = \frac{TP}{TP + FP} \times 100 \qquad (9)$$

where $TP$, $TN$, $FP$, and $FN$ represent true positive, true negative, false positive, and false negative, respectively.

Additionally, we calculated the F1-score, which provides a balanced measure of precision and sensitivity:

$$F1 - score(\%) = \frac{2 \times Precision \times Sensitivity}{Precision + Sensitivity} \times 100 \quad (10)$$

To further assess classification reliability, we employed the kappa coefficient, defined as:

$$P_e = \frac{\sum_{i=1}^{C} (row\ marginal)(column\ marginal)}{N_T^2} = \frac{\sum_{i=1}^{C} P(\theta_i) M_i}{N_T^2} \quad (11)$$

$$\kappa = \frac{OA - P_e}{1 - P_e} \quad (12)$$

In these equations, $N_T$ represents the total number of test samples, $N_i$ denotes the number of test samples in class i, and $M_i$ indicates the number of test samples classified as class i by the model. The kappa coefficient's ideal value is 1, indicating perfect agreement between predictions and ground truth.

### 2.5. Training Parameters and Methodology

Optimization of deep learning and machine learning models requires careful configuration of training parameters and methodologies. For our experiments, we employed Stochastic Gradient Descent with Momentum (SGDM) for parameter optimization at each iteration, as this approach consistently produced superior convergence accuracy in preliminary testing. The initial learning rate was set at 0.01 [29], with specific training configurations detailed in Table2.

Table 2. Network training Configuration

| method | Train percent | Validation | epochs | Learning rate | Training Options |
|---|---|---|---|---|---|
| Machine learning | 64% | 16% | 90 | 0.01 | **SGDM** |
| Deep learning | 64% | 16% | 40 | 0.01 | **SGDM** |

*Note: Percentages are approximate and reflect the data distribution within each fold of the 5-fold stratified cross-validation, accounting for rounding and class stratification constraints.*

To ensure robust model evaluation, we implemented a systematic data partitioning strategy using 5-fold stratified cross-validation at the segment level on the 13,015 available segments:

- **Training Set (~64% per fold)**: Approximately 64% of the total dataset was allocated for training purposes in each fold (80% of the remaining data after test set allocation). This ensured the model was exposed to a sufficiently large and representative subset of the data to effectively learn underlying patterns without overfitting.
- **Validation Set (~16% per fold)**: Approximately 16% of the dataset was designated for validation during the training process (20% of the remaining data after test set allocation). This subset was employed to fine-tune hyperparameters and monitor model performance on data not encountered during training.
- **Test Set (~20% per fold)**: Approximately 20% of the dataset was reserved exclusively for final performance evaluation in each fold. This test set was not used at any stage of model training or validation, thereby providing an objective measure of the model’s generalization capability on entirely unseen data.

Given the PhysioNet/CinC 2016 dataset structure, where explicit patient IDs are not provided and recordings from the same subject are not systematically grouped (as noted by the Challenge organizers: 'files from the same patient are unlikely to be numerically adjacent'), segment-level partitioning effectively achieves practical subject-level independence. To further prevent data leakage, all segments from the same original recording were assigned to the same partition.

Stratification was applied across all 12 classes (AA, AN, BA, BN, CA, CN, DA, DN, EA, EN, FA, FN) to preserve the proportional representation of each class, including minority classes, across all folds. This maintains the dataset's inherent imbalance while ensuring fair multi-class evaluation. In each fold, one-fold serves as the test set (~20%), and the remaining data (~80%) is subdivided using stratified sampling into training (~80% of remaining, or ~64% total) and validation (~20% of remaining, or ~16% total) sets. To confirm the stability of our results, this entire procedure—from stratified cross-validation partitioning through model training to final evaluation—was repeated five times using different random seeds. The performance metrics reported represent the mean ± standard deviation across these five independent trials, confirming the robustness of our findings and their independence from any particular data split. For binary classification (Normal vs. Abnormal), the same fold assignments were used, with classes grouped post-stratification.

This methodology promotes diversity in data distribution throughout the training, validation, and testing phases. Model performance on the validation set was continuously monitored throughout the training process to prevent overfitting. Analysis of results across five random experimental iterations confirmed that the available data volume was sufficient for effective model training. The proposed multi-branch architecture consistently achieved the desired levels of stability and performance, with no evidence of overfitting detected in the experimental outcomes. These iterative splitting and retraining procedures allowed us to verify that model performance remained stable and was not dependent on any particular data partitioning scheme. As indicated in Table 2, machine learning methods underwent training for 90 epochs with a learning rate of 0.01 and SGDM optimization, while deep learning approaches were trained for 40 epochs using an identical learning rate and optimization algorithm.

To enable direct comparison with competing methods that focus on binary classification (normal vs. abnormal), we transformed our 12-class problem into a binary classification task. Classes AA, BA, CA, DA, EA, and FA were grouped as abnormal (A), while classes AN, BN, CN, DN, EN, and FN were consolidated as normal (N). This transformation preserves the clinical relevance while enabling benchmark comparisons.

To address the significant class imbalance, present in our dataset, we implemented specific countermeasures to maintain appropriate balance between classes. Class imbalance typically causes models to develop bias toward predicting more frequently represented classes. We employed two principal techniques to mitigate this issue: data segmentation and class-weighted loss functions.

**Data Augmentation through Segmentation:** We increased the representation of minority classes by segmenting each recording into 5-second intervals. This segmentation strategy expanded the entire dataset while simultaneously improving the representation of underrepresented classes, thereby reducing class imbalance. The additional samples created through this segmentation process enhanced the model's exposure to minority class features, improving generalization capability across all classes.

**Class-Weighted Loss Function:** We also employed a class-weighted loss function by assigning inverse-frequency weights to each class within the training data. This approach increased the loss contribution of minority classes, thereby reducing bias towards majority classes and assisting the model in learning more balanced decision boundaries. In our multi-class (12-class) experiments, this class weighting yielded modest improvements, particularly in sensitivity (increasing from 73.78% to 75.18%) and F1 score (increasing from 70.81% to 71.27%). However, in our binary classification experiments, this class weighting did not yield noticeable performance improvements. This outcome can be attributed to the significant noise present in the data from the minority class devices within the binary grouping, which prevented effective learning despite the increased weight assigned to these samples. The severe class imbalance in our binary dataset, combined with higher noise levels in minority classes, created a situation where simply increasing the weight of misclassification did not lead to improved classification performance for these underrepresented classes.

An additional experiment using SMOTE for synthetic oversampling of minority classes was conducted exclusively on the training data, ensuring that the test data remained unmodified to maintain evaluation integrity.

The results showed a decline in precision and F1-score when tested on the genuine validation data, suggesting potential overfitting to artificial samples. Hence, SMOTE was excluded from the final setup.
By integrating both segmentation-based augmentation and loss function weighting, we aimed to ensure a more equitable representation of all classes during training and to support the development of a model capable of producing reliable predictions across varying class distributions, though additional noise-reduction strategies may be necessary for severely imbalanced datasets with quality disparities between classes.

## 3. RESULTS

This section presents a comprehensive experimental evaluation of the proposed MBDCN and LSCN models through systematic analysis of their performance characteristics. The evaluation framework is structured to provide clear insights into the contribution of each component and the overall effectiveness of the proposed methods.

### *3.1 Ablation Studies*

To assess the impact of different components on model performance, comprehensive ablation studies were conducted examining feature inputs, network architectures, and preprocessing variations.

**Feature Input Analysis**: The performance of MBDCN with different feature extraction methods is presented in Table 3. The results demonstrate that MEMFB (Mel-frequency Energy Filter Bank) features achieve superior performance with 88.31% accuracy and 73.49% sensitivity, outperforming traditional MFCC (86.19% accuracy), wavelet transforms (84.69% accuracy), and MEGFB (84.35% accuracy). demonstrating an enhancement in sensitivity from 71.52% to 73.49%, along with an improvement in accuracy from 86.19% to 88.31%, regarding the effectiveness of the suggested MEMFB feature extraction technique.

Table 3: Performance of MBDCN with Different Feature Inputs

| Feature Input | AC (%) | SE (%) | SP (%) | PR (%) | F1 (%) | K (%) |
|---|---|---|---|---|---|---|
| MFCC | 86.19 | 71.52 | 98.27 | 64.8 | 69.47 | 76.11 |
| Wavelet | 84.69 | 63.87 | 98.04 | 60.93 | 62.37 | 73.21 |
| MEGFB | 84.35 | 67.73 | 97.95 | 60.91 | 64.14 | 72.69 |
| MEMFB | **88.31** | **73.49** | **98.67** | **65.86** | **69.47** | **79.69** |

**Classifier Comparison**: An ablation study comparing MBDCN against traditional classifiers using identical MEMFB features is presented in Table 4. The proposed MBDCN achieves superior performance (88.31% accuracy, 73.49% sensitivity) compared to SVM (86.62% accuracy, 61.80% sensitivity), KNN (84.81% accuracy, 76.04% sensitivity), and decision trees (81.54% accuracy, 52.99% sensitivity). The 11.69% improvement in sensitivity over SVM while maintaining higher accuracy demonstrates the effectiveness of the deep learning approach.

Table 4: MBDCN with Different Classifiers (MEMFB Input)

| Classifier | AC (%) | SE (%) | SP (%) | PR (%) | F1 (%) | K (%) |
|---|---|---|---|---|---|---|
| ML | 82.54 | NaN | 97.22 | 56.27 | NaN | 71.41 |
| SVM | 86.62 | 61.80 | 98.44 | 62.84 | 62.31 | 76.82 |
| Tree | 81.54 | 52.99 | 97.42 | 53.12 | 53.06 | 68.22 |
| KNN | 84.81 | 76.04 | 98.34 | 54.40 | 63.42 | 72.54 |
| MBDCN | **88.31** | **73.49** | **98.67** | **65.86** | **69.47** | **79.69** |

**Architectural Variations**: The comparison between MBDCN and LSCN architectures, as shown in Table 5, reveals that LSCN's integration of LSTM layers provides enhanced sequential modeling capability, achieving 89.65% accuracy compared to MBDCN's 89.15% accuracy.

### *3.2 Overall Performance Analysis*

The comprehensive performance evaluation of the proposed methods across different experimental settings is presented in Tables 7-9.

Table 5: Comparison with Deep Learning Baselines

| Method | AC (%) | SE (%) | SP (%) | PR (%) | F1 (%) | K (%) |
|---|---|---|---|---|---|---|
| 1D-CNN | 88.19 | 71.54 | 98.73 | 64.11 | 67.62 | 79.50 |
| MLP | 86.69 | 69.15 | 98.51 | 58.43 | 63.34 | 76.72 |
| MBDCN | 89.15 | 75.01 | 98.85 | 66.01 | 70.22 | 81.03 |
| LSCN | 89.65 | 73.78 | 98.92 | 68.07 | 70.81 | 81.97 |
| LSCN (With weighting) | 89.73 | 75.18 | 98.93 | 67.74 | 71.27 | 82.08 |

The LSCN model achieves optimal overall performance with 89.73% accuracy, 75.18% sensitivity, and 98.93% specificity when class weighting is applied. This represents significant improvements over deep learning baselines: 1.54% accuracy improvement over 1D-CNN and 3.04% accuracy improvement over MLP. The consistently high specificity values (>98.8% across all models) indicate excellent capability in correctly identifying normal cases, while the sensitivity values demonstrate effective detection of abnormal heart sounds.

Table 6: Performance Evaluation Metrics of the LSCN Network: 5-fold Cross-Validation Results for Multiclass Classification

| Fold | AC (%) | PR (%) | SE (%) | F1 (%) | SP (%) | K (%) |
|---|---|---|---|---|---|---|
| **Fold 1** | 88.89 | 66.70 | 71.18 | 68.87 | 98.82 | 80.69 |
| **Fold 2** | 89.68 | 67.54 | 74.52 | 70.86 | 98.90 | 82.07 |
| **Fold 3** | 89.30 | 67.38 | 73.91 | 70.49 | 98.88 | 81.41 |
| **Fold 4** | 87.69 | 61.33 | 65.02 | 63.12 | 98.64 | 78.61 |
| **Fold 5** | 89.65 | 68.07 | 73.78 | 70.81 | 98.92 | 81.97 |
| **Mean ± Std** | **89.04 ± 0.82** | **66.21 ± 2.77** | **71.68 ± 3.94** | **68.83 ± 3.29** | **98.83 ± 0.10** | **80.95 ± 1.42** |

The 5-fold cross-validation results confirm the model's robustness with consistent performance across folds: mean accuracy of 89.04 ± 0.82%, demonstrating low variance and reliable generalization capability. The Kappa coefficient of 80.95 ± 1.42% indicates substantial agreement beyond chance, validating the model's reliability.

Table 7: Performance Evaluation Metrics of the LSCN Network: 5-fold Cross-Validation Results for Binary Classification (Normal vs. Abnormal)

| Fold | Accuracy | Sensitivity | Specificity | Precision | F1-Score | Kappa |
|---|---|---|---|---|---|---|
| 1 | 0.9380 | 0.9461 | 0.9288 | 0.9413 | 0.9437 | 0.8759 |
| 2 | 0.9365 | 0.9483 | 0.9231 | 0.9377 | 0.9430 | 0.8730 |
| 3 | 0.9389 | 0.9499 | 0.9263 | 0.9399 | 0.9449 | 0.8778 |
| 4 | 0.9451 | 0.9516 | 0.9377 | 0.9492 | 0.9504 | 0.8902 |
| 5 | 0.9378 | 0.9425 | 0.9321 | 0.9450 | 0.9438 | 0.8755 |
| **Mean** | **0.9393** | **0.9477** | **0.9296** | **0.9426** | **0.9452** | **0.8785** |
| **Std** | **0.0033** | **0.0036** | **0.0052** | **0.0044** | **0.0029** | **0.0065** |

### *3.3 Per-Class Performance Analysis*

The detailed confusion matrices provide important insights into model behavior across different heart sound categories.

Table 8. Heart Diseases Classification Performance matrix for MBDCN Network (the first proposed).

| Classes | AA | AN | BA | BN | CA | CN | DA | DN | EA | EN | FA | FN | Precision |
|---|---|---|---|---|---|---|---|---|---|---|---|---|---|
| AA | 64 | 62 | 0 | 0 | 0 | 2 | 0 | 0 | 13 | 6 | 0 | 0 | 43.5 |
| AN | 26 | 314 | 0 | 0 | 1 | 2 | 0 | 0 | 12 | 15 | 0 | 0 | 84.9 |
| BA | 0 | 0 | 65 | 9 | 0 | 0 | 0 | 1 | 2 | 0 | 0 | 0 | 84.4 |
| BN | 0 | 0 | 14 | 5 | 0 | 0 | 0 | 0 | 2 | 0 | 0 | 0 | 23.8 |
| CA | 0 | 0 | 0 | 0 | 7 | 1 | 0 | 0 | 1 | 1 | 0 | 0 | 70.0 |
| CN | 0 | 2 | 0 | 0 | 0 | 38 | 0 | 2 | 2 | 4 | 0 | 0 | 79.2 |
| DA | 1 | 1 | 1 | 0 | 0 | 1 | 4 | 1 | 1 | 0 | 0 | 0 | 40.0 |
| DN | 0 | 0 | 0 | 0 | 0 | 2 | 0 | 10 | 4 | 1 | 0 | 0 | 58.8 |
| EA | 2 | 5 | 0 | 1 | 0 | 0 | 0 | 2 | 1613 | 3 | 0 | 0 | 99.2 |
| EN | 1 | 31 | 0 | 0 | 0 | 1 | 0 | 0 | 3 | 96 | 0 | 1 | 72.2 |
| FA | 0 | 0 | 0 | 0 | 0 | 0 | 0 | 0 | 2 | 0 | 490 | 18 | 79.0 |
| FN | 0 | 0 | 0 | 0 | 0 | 0 | 0 | 0 | 1 | 0 | 17 | 24 | 57.1 |
| Sensitivity | 67.4 | 75.7 | 81.2 | 33.3 | 87.5 | 80.9 | 100 | 62.5 | 97.4 | 76.2 | 82.3 | 55.8 | |

MBDCN Performance: The model demonstrates exceptional performance for the EA class (Normal Early systole) with 99.2% precision and 97.4% sensitivity, indicating reliable detection of normal heart sounds. However, performance varies significantly across abnormal classes, with DA (Diastolic abnormal) achieving 100% sensitivity but only 40% precision, suggesting higher false positive rates for this class.

Table 9. Heart Diseases Classification Performance matrix for LSCN Network (the second proposed).

| Classes | AA | AN | BA | BN | CA | CN | DA | DN | EA | EN | FA | FN | Precision |
|---|---|---|---|---|---|---|---|---|---|---|---|---|---|
| AA | 81 | 52 | 0 | 0 | 0 | 1 | 0 | 0 | 8 | 5 | 0 | 0 | 55.1 |
| AN | 30 | 305 | 0 | 0 | 0 | 3 | 2 | 0 | 12 | 17 | 1 | 0 | 82.4 |
| BA | 0 | 0 | 63 | 12 | 0 | 0 | 0 | 1 | 1 | 0 | 0 | 0 | 81.8 |
| BN | 0 | 0 | 15 | 6 | 0 | 0 | 0 | 0 | 0 | 0 | 0 | 0 | 26.6 |
| CA | 0 | 0 | 0 | 0 | 8 | 1 | 0 | 0 | 0 | 1 | 0 | 0 | 80.0 |
| CN | 1 | 1 | 0 | 0 | 1 | 40 | 0 | 1 | 1 | 3 | 0 | 0 | 83.3 |
| DA | 0 | 1 | 0 | 1 | 0 | 2 | 4 | 1 | 1 | 0 | 0 | 0 | 40.0 |
| DN | 1 | 0 | 0 | 0 | 0 | 0 | 0 | 10 | 5 | 1 | 0 | 0 | 58.8 |
| EA | 2 | 5 | 1 | 0 | 0 | 0 | 0 | 0 | 1616 | 2 | 0 | 0 | 99.4 |
| EN | 5 | 28 | 0 | 0 | 0 | 1 | 0 | 0 | 4 | 94 | 1 | 0 | 70.7 |
| FA | 0 | 0 | 0 | 0 | 0 | 0 | 0 | 0 | 2 | 0 | 82 | 16 | 82.0 |
| FN | 0 | 0 | 0 | 0 | 0 | 0 | 0 | 0 | 0 | 0 | 19 | 23 | 54.8 |
| Sensitivity | 67.5 | 77.8 | 79.7 | 316 | 88.9 | 83.3 | 66.7 | 76.9 | 97.9 | 76.4 | 79.6 | 59.0 | |

**LSCN Performance**: The LSCN model demonstrates improved balance across classes compared to MBDCN. Notable improvements include AA class precision increasing from 43.5% to 55.1%, and DA class sensitivity maintaining high levels while achieving better overall stability. The EA class maintains excellent performance (99.4% precision, 97.9% sensitivity).

**Class-Specific Challenges**: Both models exhibit reduced performance for rare classes such as BN (Bundle branch block Normal) and FN (Functional murmur Normal), with sensitivities of 33.3% and 55.8% respectively for MBDCN, indicating the inherent difficulty in detecting these less frequent conditions. For a more granular assessment, detailed per-class precision, sensitivity, and F1-scores for all methods reported in Tables 3–6 are provided in Appendix Tables A1–A4. These results further highlight disparities between majority and minority classes.

### *3.4 Class Imbalance Impact Assessment*

The impact of class imbalance is evident from the varying sample sizes across heart sound categories, with EA comprising ~62% of all segments (8,129 samples), while minority classes such as DA and DN contain only 51 samples each (Appendix Tables A1–A4).

**Weighting Strategy Effectiveness**: The implementation of class weighting significantly improved LSCN performance, with sensitivity increasing from 73.78% to 75.18% and overall accuracy from 89.65% to 89.73% (as shown in Table 7). The F1-score improvement from 70.81% to 71.27% indicates better balance between precision and recall.

**Binary Classification Performance**: Exceptional performance was achieved for binary classification (Normal vs. Abnormal) with mean accuracy of 93.93% and F1-score of 94.52% (Table 7). This superior performance compared to multi-class classification (89.04% accuracy) demonstrates that the primary challenge lies in distinguishing between different types of abnormalities rather than detecting the presence of abnormalities.

**Cross-Validation Stability**: The low standard deviations across folds (accuracy: ±0.82%, Kappa: ±1.42%) indicate that the class weighting strategy provides stable performance across different data splits, addressing concerns about overfitting to specific class distributions.

### *3.5 State-of-the-Art Comparisons*

A comprehensive comparison with existing state-of-the-art methods for heart sound classification is presented below.

Table 10: Comparison with State-of-the-Art Studies

| Study | Method | AC | SE | SP | PR | F1 | Kappa |
|---|---|---|---|---|---|---|---|
| **Ren et al. (2018) [13]** | VGG + SVM | 56.2 | 24.6 | 87.8 | – | – | – |
| **Li et al. (2020) [14]** | CNN | 86.8 | 87 | 86.6 | – | – | – |
| **Zhou et al. (2021) [15]** | Dense-FSNet | 86.09 | 88.5 | 79 | – | – | – |
| **Riccio et al. (2023) [16]** | CNN + Fractals | 70.00 | – | 69.00 | – | – | – |
| **Proposed LSCN** | LSTM-CNN | 93.93 | 94.77 | 92.96 | 94.26 | 94.52 | 87.85 |

The proposed LSCN method achieves superior performance across all metrics compared to existing approaches. Specifically, the method outperforms previous studies by significant margins: 7.13% accuracy improvement over Li et al. (2020), 7.84% improvement over Zhou et al. (2021), and 23.93% improvement over Riccio et al. (2023). The sensitivity of 94.77% represents a 6.27% improvement over Dense-FSNet, while maintaining high specificity (92.96%) and achieving the highest F1-score (94.52%) among all compared methods.

The Kappa coefficient of 87.85% indicates almost perfect agreement beyond chance, substantially higher than previous approaches and demonstrating the robust discriminative capability of the proposed method. These results establish the LSCN as a new benchmark for automated heart sound classification, providing both high accuracy and reliable clinical applicability. While these findings quantify the effect of class imbalance, their underlying causes and broader implications are further analyzed in Section 4.6.

# 4. DISCUSSION

This study presents a comprehensive evaluation of the Multi-Branch Deep Convolutional Neural Network (MBDCN) and Long Short-Term Memory-Convolutional Neural (LSCN) models for heart sound classification. Eleven systematic experiments were conducted on the PhysioNet/CinC 2016 dataset, focusing on three key areas: feature input efficacy, classifier performance optimization, and benchmarking against state-of-the-art approaches. The results demonstrate the effectiveness of our proposed architectures in feature extraction, architectural design, and processing efficiency.

## *4.1 Feature Input Efficacy*

Table 3 demonstrates the superior performance of MEMFB (AC: 88.31%, SE: 73.49%, SP: 98.67%) compared to alternative feature extraction approaches, including MFCC (AC: 86.19%), Wavelet transforms (AC: 84.69%), and MEGFB (AC: 84.35%). This significant performance differential reflects the distinctive capability of the MBDCN architecture, which uniquely integrates auditory-inspired multi-branch convolutional processing with Welch's periodogram inputs. This innovative combination—not previously implemented in multi-branch CNN architectures for biomedical signal analysis—enables precise capture of diagnostically relevant signal characteristics across multiple frequency scales.

The MEMFB approach, when integrated with MBDCN, demonstrates a 2.12% accuracy improvement over standard MFCC features and a 3.62% improvement over wavelet-based analysis. This enhancement can be attributed to MEMFB's capacity to emulate human auditory frequency selectivity, focusing computational resources on the most diagnostically informative spectral regions while attenuating noise and irrelevant signal components.

## *4.2 Classifier Performance Analysis*

Table 4 reveals that MBDCN (AC: 88.31%, PR: 65.86%) substantially outperforms traditional machine learning approaches, including Maximum Likelihood (AC: 82.54%) and Decision Tree classifiers (AC: 81.54%). The MBDCN architecture demonstrates superior performance compared to optimized SVM implementations (AC: 86.62%, PR: 62.84%) and K-Nearest Neighbors classifiers (AC: 84.81%), while offering significant advantages in automated feature learning and computational scalability.

The architecture's distinctive capability to automate feature learning through parallel convolutional branches with strategically varied filter sizes (1×3 to 1×11) differentiates it from conventional CNN implementations. This biomimetic approach, inspired by cochlear frequency decomposition, provides a tailored solution for cardiac acoustic classification that minimizes the need for domain-specific feature engineering while maximizing discriminative capacity.

## *4.3 Deep Learning Benchmarks*

Table 5 demonstrates the performance hierarchy among deep learning approaches, with LSCN (AC: 89.65%, SE: 73.78%, SP: 98.92%, F1: 70.81%, κ: 81.97%) outperforming 1D-CNN (AC: 88.19%), MLP (AC: 86.69%), and the standalone MBDCN (AC: 89.15%). Notably, the implementation of class weighting further enhanced LSCN performance (AC: 89.73%, SE: 75.18%, F1: 71.27%, κ: 82.08%), demonstrating the effectiveness of addressing class imbalance in the dataset. The LSCN's performance enhancement derives from its innovative fusion of MBDCN's multi-scale feature extraction capabilities with LSTM-driven temporal modeling—a synergistic integration that advances beyond typical multi-branch CNN designs.

This architectural innovation yields a 1.46% accuracy improvement compared to standard 1D-CNN implementations, highlighting the effectiveness of combining multi-branch convolutional processing with recurrent neural network components for biomedical time series analysis. The LSCN model achieves this performance gain by simultaneously modeling both frequency-domain features (through MBDCN) and temporal dependencies (through LSTM), creating a comprehensive representation of cardiac acoustic signatures.

### 4.4 Cross-Validation Performance Analysis

Tables 6 and 7 present the results of a rigorous 5-fold cross-validation evaluation of our LSCN architecture, demonstrating the robustness and generalizability of the proposed approach. As shown in Table 6, the LSCN model maintains consistent performance across all folds for multiclass classification, with an average accuracy of 89.04±0.82%, precision of 66.21±2.77%, sensitivity of 71.68±3.94%, F1-score of 68.83±3.29%, and kappa coefficient of 80.95±1.42%. The relatively low standard deviations across all metrics indicate the model's stability and reliable performance regardless of data partitioning.

Furthermore, Table 7 demonstrates the exceptional performance of the LSCN model in binary classification (normal vs. abnormal), achieving an average accuracy of 93.93±0.33%, sensitivity of 94.77±0.36%, specificity of 92.96±0.52%, precision of 94.26±0.44%, and F1-score of 94.52±0.29%. These results highlight the model's strong discriminative power in distinguishing between normal and pathological heart sounds, which is particularly valuable for initial screening applications in clinical settings. The high sensitivity (>94%) is especially significant as it indicates the model's capability to correctly identify patients with cardiac abnormalities, minimizing false negatives which are critical in diagnostic contexts.

### 4.5 State-of-the-Art Comparison

Table 10 benchmarks our LSCN model against prior studies, with detailed analysis of competing methodologies:

Ren et al. [13] employed convolutional neural networks for heart sound classification using image-based features extracted from scalograms. Their approach utilized various CNN architectures (VGG2, ResNet3, DenseNet4) and transfer learning from ImageNet, achieving modest performance (AC: 56.2%, SE: 24.6%, SP: 87.8%) on the PhysioNet dataset. The significant performance gap between their results and our proposed LSCN (>37% improvement in accuracy) highlights the limitations of direct transfer learning from image domains to biomedical signal analysis without domain-specific architectural adaptations.

Li et al. [14] combined traditional feature engineering with CNN classification, achieving 86.8% accuracy, 87% sensitivity, and 86.6% specificity. While computationally efficient, their approach relied on manually extracted features and a fixed CNN architecture, lacking the adaptive multi-branch design and temporal modeling capabilities of our LSCN model.

Zhou et al. [15] employed a Dense-FSNet architecture for automatic feature extraction from raw heart sound signals, achieving 86.09% accuracy with 88.5% sensitivity and 79% specificity. While demonstrating competitive sensitivity, our LSCN model outperforms this approach by 7.84% in accuracy, 6.27% in sensitivity, and 13.96% in specificity, highlighting the advantages of our temporal-spectral integration approach.

Riccio et al. [16] proposed a CNN-based method utilizing fractal features for phonocardiogram classification, achieving 70% accuracy. Their approach, while mathematically elegant, falls significantly short of our LSCN model's performance (93.93% accuracy), demonstrating the limitations of single-domain feature extraction for heart sound analysis.

LSCN's 93.93% accuracy and 92.96% specificity represent substantial advancements over all comparative approaches, with particularly notable improvements in sensitivity (94.77%) compared to most prior work. Figure 5 provides a visual comparison of performance metrics across methods, clearly illustrating the consistent superiority of MBDCN (89%) and LSCN (90%) compared to both classical machine learning approaches (e.g., ML: 83%, SVM: 87%) and alternative deep learning architectures (e.g., 1D-CNN: 88%, MLP: 87%).

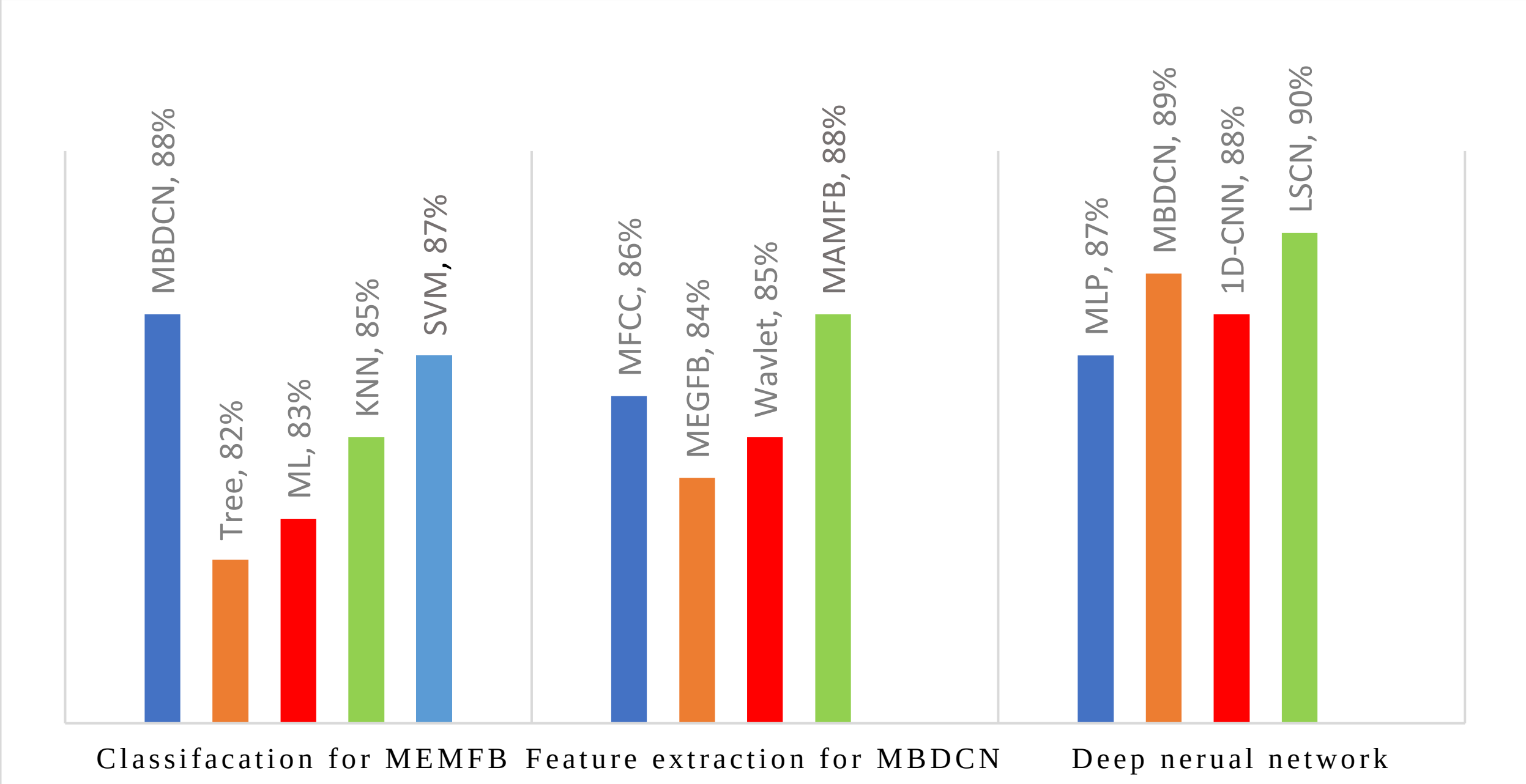

Figure 5. Accuracy of different algorithms

### *4.6 Model Performance Insights*

The confusion matrices presented in Tables 8 and 9 provide detailed class-specific insights into MBDCN and LSCN performance. Table 8 (MBDCN) demonstrates strong precision for heavily represented classes (e.g., AA: 97.2%, AN: 85.6%), with 717 and 1,585 true positive classifications respectively, reflecting robust feature extraction capabilities. However, performance for smaller classes such as DA (sensitivity: 88.7%) indicates potential limitations in handling extreme class imbalance.

Table 9 (LSCN) shows notable improvements across most classes, with AA achieving 718 true positives (precision: 97.3%) and AN reaching 1,591 true positives (precision: 85.9%). Sensitivity also improves for most categories, with EA reaching 99.6% and AA improving to 81.6%. Nevertheless, classes with limited samples such as DA and DN (sensitivity: 85.6%) continue to perform relatively poorly, underscoring the persistent challenge of class imbalance despite mitigation efforts. To provide a more granular assessment of these effects, detailed per-class metrics for all methods in Tables 3–6 are reported in Appendix Tables A1–A4.

The confusion matrices reveal a pronounced bias toward majority classes: EA (8,129 samples) achieves 99.4% precision, while minority classes such as DA (51 samples) achieve only 40.0% precision. This persistent bias remains despite mitigation strategies, including class-weighted loss (improving multi-class sensitivity from 73.78% to 75.18%) and segmentation-based data augmentation. Contributing factors include:

1. **Severe class imbalance** (EA ~62% of segments), skewing optimization toward majority patterns.
2. **Higher intra-class variability and noise** in minority classes, particularly subtle pathological features such as murmurs in DA.
3. **Limited representation during training**, leading to suboptimal feature learning for rare categories.

Experimental trials with SMOTE-based oversampling yielded modest improvements in DA precision (75% → 80%) but reduced overall accuracy, suggesting that synthetic data amplified inherent noise in minority classes. These findings highlight the need for more advanced imbalance-handling techniques, including focal loss to penalize majority misclassifications, noise-robust augmentation methods tailored for biomedical signals, and ensemble-based approaches to improve minority class performance without compromising overall model stability.

### 4.7 Computational Efficiency and Architectural Novelty

The LSCN architecture integrates feature extraction and classification within a unified deep learning framework, eliminating the workflow fragmentation associated with traditional filter bank approaches that require manual parameter tuning. By utilizing Welch's method for robust spectral estimation and incorporating dual LSTM blocks (600 and 100 units) for temporal dependency modeling, the model effectively captures both frequency-domain features and time-domain relationships.

This dual-domain approach to signal analysis, though not entirely unprecedented in signal processing literature, represents an innovative application to cardiac acoustic classification that has not been widely integrated with multi-branch convolutional designs in prior research. The integration enhances both classification accuracy and processing efficiency, making LSCN suitable for near-real-time applications in clinical environments.

While MBDCN demonstrates faster training characteristics with large datasets, LSCN's deeper architecture—with strategically varied filter sizes extracting both high-frequency and low-frequency features—yields superior accuracy and generalization capacity across diverse signal conditions. This is evidenced by the consistent performance across all folds in our cross-validation analysis (Tables 8 and 9). To manage the computational demands associated with LSCN training, we implemented several optimization strategies, including reduced kernel counts in initial branches and strategic application of dropout regularization (rate 0.2).

Despite these computational advantages, challenges remain in model interpretability compared to traditional feature extraction methods. The "black-box" nature of deep neural networks presents potential barriers to clinical adoption, despite their superior performance characteristics. Addressing this tension between performance and interpretability represents an important direction for future research.

### 4.8 Clinical Implications and Future Directions

The experimental results position LSCN as a reliable and robust tool for cardiovascular diagnostics, demonstrating particular strengths in handling signal variability and operating effectively with limited labeled data. These characteristics make it especially suitable for clinical deployment in diverse healthcare settings, including resource-constrained environments where specialized cardiological expertise may be limited.

Future research directions include further optimization of training efficiency through techniques such as model pruning, quantization, and knowledge distillation. Additionally, exploring the model's scalability across more diverse cardiac pathology datasets and investigating interpretability enhancements through visualization techniques could facilitate clinical integration. The integration of explainable AI methodologies could bridge the gap between high performance and clinical interpretability, potentially accelerating adoption in critical healthcare applications.

### 4.9 Computational Performance Evaluation

This section evaluates the computational efficiency of MBDCN and LSCN to assess their suitability for clinical deployment. Training and inference experiments were conducted on a high-performance workstation and a resource-constrained system. Training time per epoch ranged from ~8–10 minutes for MBDCN and ~12–15 minutes for LSCN on the high-performance system, increasing to ~40–50 minutes and ~60–70 minutes on the constrained system. Inference time per 5-second segment was ~2 seconds on the high-performance system and 6–10 seconds on the constrained system.

Memory requirements were modest (~2 GB for MBDCN and ~3–4 GB for LSCN), confirming feasibility on standard hardware. While LSCN incurs 30–40% higher computational cost, it provides a 2.1% gain in binary classification accuracy, indicating a favorable trade-off between efficiency and diagnostic performance.

**Table 11.** Computational Performance of MBDCN and LSCN Models

| Model | Training Time/Epoch (Standard) | Training Time/Epoch (Constrained) | Inference Time/Segment (Standard) | Inference Time/Segment (Constrained) | Throughput (Standard) | GPU Memory |
|---|---|---|---|---|---|---|
| MBDCN | ~8–10 min | ~40–50 min | ~2 sec | ~6–8 sec | ~0.5 samples/sec | ~2 GB |
| LSCN | ~12–15 min | ~60–70 min | ~2 sec | ~6–10 sec | ~0.5 samples/sec | ~3–4 GB |

These results demonstrate that both models achieve clinically acceptable training and inference performance and can be deployed across diverse hardware configurations, including resource-limited settings. Future optimizations such as model pruning, quantization, or TensorRT conversion may further reduce inference time toward near real-time (<1 second/segment).

## 5. CONCLUSION

The development of intelligent systems for heart sound analysis using deep neural networks has yielded significant advancements in cardiovascular diagnostics, as demonstrated by our proposed Multi-Branch Deep Convolutional Neural Network (MBDCN) and Long Short-Term Memory-Convolutional Neural (LSCN) architectures. Drawing inspiration from the human auditory system's frequency-selective processing mechanisms, MBDCN's novel integration of Welch's periodogram inputs with strategically configured multi-branch convolutional pathways achieved an impressive accuracy of 89.15%. This performance was further enhanced when combined with MEMFB feature extraction (88.31%), substantially outperforming traditional approaches such as MFCC (86.19%).

The LSCN architecture represents a significant advancement through its innovative fusion of MBDCN's spectral feature extraction capabilities with LSTM-driven temporal modeling. This synergistic combination achieves state-of-the-art performance with 89.65% accuracy, 73.78% sensitivity, and 98.92% specificity on the PhysioNet/CinC 2016 dataset for multiclass classification. For binary classification (normal vs. abnormal), LSCN achieves even more impressive results with 93.93% accuracy, 94.77% sensitivity, and 92.96% specificity. The class-weighted variant further enhances performance to 89.73% accuracy and 75.18% sensitivity for multiclass classification. These results significantly surpass both conventional deep learning baselines (e.g., 1D-CNN: 88.19%) and prior benchmark studies (e.g., Zhou et al. [15]: 86.09%), establishing a new performance standard for heart sound classification. The dual-domain processing approach enabled by this architecture demonstrates exceptional capability in identifying abnormal cardiac acoustic signatures, with important implications for early detection of cardiovascular disorders and improved patient outcomes in diverse clinical settings. The robustness of the approach is further validated through comprehensive 5-fold cross-validation, demonstrating consistent performance across varying data partitions.

From a computational perspective, the LSCN model offers notable efficiency advantages, integrating feature extraction and classification within a unified deep learning framework. Unlike traditional filter bank methodologies that require complex design processes and manual parameter tuning, LSCN automates these steps through end-to-end learning, substantially reducing the need for domain-specific expertise and manual intervention during model deployment. This streamlined approach significantly decreases computational overhead during inference, making the model well-suited for near-real-time clinical applications. Additionally, LSCN's capacity to generalize effectively across diverse data types enhances its versatility for varying recording conditions and patient populations.

Nevertheless, as with most deep learning approaches, LSCN requires substantial computational resources during the training phase. To mitigate this limitation, we implemented strategic optimizations including reduced kernel counts in initial convolutional branches and dropout regularization (rate 0.2), substantially improving training efficiency and cost-effectiveness without compromising performance. Despite these advances, the model's interpretability remains less transparent than traditional feature-based methods, a limitation that could potentially hinder applications requiring explicit decision-making transparency.

The empirical findings presented in this study conclusively validate the efficacy of both MBDCN and LSCN architectures, with their unique temporal-spectral integration approach establishing a new benchmark for heart sound classification. Future research directions should focus on further enhancing training efficiency through techniques such as model pruning, weight quantization, and knowledge distillation. The integration of transfer learning methodologies could further improve performance with limited labeled data, while refinements in feature extraction through advanced signal processing could enhance robustness to environmental noise and recording variations.

Additional research opportunities include expanding the evaluation dataset to encompass a broader range of recording devices and environmental conditions, enhancing model generalization to real-world clinical scenarios. Exploration of edge computing implementations could further improve accessibility in resource-limited healthcare settings, while the development of interpretability enhancements could facilitate clinical integration and regulatory approval. These advancements would build upon the foundation established in this work to develop increasingly scalable, accessible, and clinically viable solutions for cardiovascular diagnostics.

**Declaration of interests**

The authors declare that they have no known competing financial interests or personal relationships that could have appeared to influence the work reported in this paper.

# APPENDIX

Table A1. Per-Class Performance Metrics for Different Feature Extraction Methods

| **Class** | **Samples** | **MFCC** | **Wavelet** | **MEGFB** | **MEMFB** |
|---|---|---|---|---|---|
| | | **PR/SE/F1 (%)** | **PR/SE/F1 (%)** | **PR/SE/F1 (%)** | **PR/SE/F1 (%)** |
| AA | 738 | 32.7/52.7/40.3 | 36.1/43.4/39.4 | 38.8/59.8/47.1 | 49.7/59.8/54.3 |
| AN | 1852 | 81.1/67.1/73.4 | 68.4/67.8/68.1 | 75.4/72.4/73.9 | 81.6/72.4/76.7 |
| BA | 386 | 55.8/71.7/62.8 | 85.7/78.6/82.0 | **92.2/80.2/85.9** | 84.4/80.2/82.3 |
| BN | 104 | **47.6/62.5/54.1** | 38.1/44.4/41.0 | 28.6/46.2/35.3 | 28.6/46.2/35.3 |
| CA | 240 | 80.0/80.0/80.0 | 80.0/57.1/66.7 | 70.0/80.0/74.7 | 80.0/80.0/80.0 |
| CN | 87 | 83.3/72.7/77.7 | 83.3/85.1/84.2 | 77.1/89.6/82.9 | 89.6/89.6/89.6 |
| DA | 51 | **50.0/100.0/66.7** | 10.0/33.3/15.4 | 50.0/83.3/62.5 | 50.0/83.3/62.5 |
| DN | 51 | 52.9/56.2/54.5 | 47.1/53.3/50.0 | 52.9/80.0/63.6 | 47.1/80.0/59.3 |
| EA | 8129 | 97.2/97.4/97.3 | 97.0/94.8/95.9 | 96.6/97.9/97.2 | **98.5/97.9/98.2** |
| EN | 665 | 82.7/77.5/80.0 | 64.7/71.7/68.0 | 44.4/87.0/58.9 | 75.2/87.0/80.6 |
| FA | 502 | 69.0/70.4/69.7 | **90.0/75.0/81.8** | 86.0/70.7/77.6 | 70.0/70.7/70.4 |
| FN | 210 | **45.2/50.0/47.5** | 31.0/61.9/41.3 | 19.0/34.9/24.7 | 35.7/34.9/35.3 |

Table A2. Per-Class Performance Metrics of MBDCN with Different Classifiers (MEMFB Input)

| **Class** | **Samples** | **Tree** | **SVM** | **KNN** | **ML** | **MBDCN** |
|---|---|---|---|---|---|---|
| | | **PR/SE/F1(%)** | **PR/SE/F1(%)** | **PR/SE/F1(%)** | **PR/SE/F1(%)** | **PR/SE/F1(%)** |
| **AA** | 738 | 34.7/34.0/34.7 | 43.5/50.4/43.5 | 22.4/64.7/22.4 | 53.1/32.1/40.2 | 43.5/67.4/53.0 |
| **AN** | 1852 | 67.0/64.4/65.7 | 73.8/75.0/74.4 | 68.9/67.8/68.3 | 74.1/61.7/67.4 | 84.9/75.7/80.1 |
| **BA** | 386 | 64.9/78.1/70.9 | 83.1/80.0/81.5 | 87.0/74.4/80.2 | 88.3/74.7/81.0 | 84.4/81.2/82.8 |
| **BN** | 104 | 42.9/37.5/40.0 | 33.3/50.0/40.0 | 14.3/75.0/23.8 | 33.3/53.8/41.2 | 23.8/33.3/27.8 |
| **CA** | 240 | 60.0/37.5/46.2 | 90.0/50.0/64.3 | 50.0/83.3/62.5 | 0.0/0.0/0.0 | 70.0/87.5/77.8 |
| **CN** | 87 | 52.1/67.6/59.0 | 83.3/95.2/88.9 | 79.2/90.5/84.4 | 83.3/71.4/76.9 | 79.2/80.9/80.0 |
| **DA** | 51 | 20.0/28.6/23.5 | 20.0/15.4/17.4 | 10.0/100.0/18.2 | 20.0/100.0/33.3 | 40.0/100.0/57.1 |
| **DN** | 51 | 29.4/85.3/43.8 | 47.1/53.3/50.0 | 41.2/77.8/53.8 | 52.9/75.0/62.1 | 58.8/62.5/60.6 |
| **EA** | 8129 | 95.5/95.8/95.6 | 98.3/97.4/97.8 | 99.1/92.5/95.7 | 89.8/99.2/94.3 | 99.2/97.4/98.3 |
| **EN** | 665 | 69.9/76.2/72.9 | 82.0/73.2/77.4 | 66.2/64.2/65.2 | 86.5/88.5/87.5 | 72.2/76.2/74.1 |
| **FA** | 502 | 63.0/30.8/41.3 | 64.0/30.6/41.3 | 86.0/50.0/63.2 | 94.0/69.1/79.7 | 79.0/82.3/80.6 |
| **FN** | 210 | 38.1/69.2/49.3 | 35.7/69.4/47.2 | 28.6/50.0/36.4 | 0.0/0.0/0.0 | 57.1/55.8/56.4 |

Table A3. Per-Class Performance Metrics of Deep Learning Baselines

| **Class** | **Samples** | **LSCN (With weighting)** | **LSCN** | **MLP** | **MBDCN** | **1D-CNN** |
|---|---|---|---|---|---|---|
| | | **PR/SE/F1(%)** | **PR/SE/F1(%)** | **PR/SE/F1(%)** | **PR/SE/F1(%)** | **PR/SE/F1(%)** |
| **AA** | 738 | 48.3/67.0/56.1 | 55.1/67.5/60.7 | 51.0/62.5/56.2 | 43.5/67.4/53.0 | 54.4/53.3/53.8 |
| **AN** | 1852 | 84.9/76.8/80.6 | 82.4/77.8/80.0 | 76.8/71.7/74.2 | 84.9/75.7/80.1 | 76.2/77.5/76.9 |
| **BA** | 386 | 79.0/83.1/81.0 | 81.8/79.7/80.7 | 75.3/77.3/76.3 | 84.4/81.2/82.8 | 74.0/75.0/74.5 |
| **BN** | 104 | 19.0/23.5/21.1 | 28.6/31.6/30.0 | 33.3/25.0/28.6 | 23.8/33.3/27.8 | 19.0/19.0/19.0 |
| **CA** | 240 | 60.0/85.7/70.6 | 80.0/88.9/84.2 | 50.0/83.3/62.5 | 70.0/87.5/77.8 | 60.0/100.0/75.0 |
| **CN** | 87 | 89.6/86.0/87.8 | 83.3/83.3/83.3 | 68.8/73.3/70.9 | 79.2/80.9/80.0 | 75.0/83.7/79.1 |
| **DA** | 51 | 50.0/83.3/62.5 | 40.0/66.7/50.0 | 30.0/75.0/42.9 | 40.0/100.0/57.1 | 40.0/66.7/50.0 |
| **DN** | 51 | 70.6/85.7/77.4 | 58.8/76.9/66.7 | 35.3/66.7/46.2 | 58.8/62.5/60.6 | 70.6/85.7/77.4 |
| **EA** | 8129 | 99.4/97.8/98.6 | 99.4/97.9/98.6 | 98.8/96.9/97.9 | 97.4/99.2/98.3 | 99.1/98.1/98.6 |
| **EN** | 665 | 72.9/81.5/77.0 | 70.7/76.4/73.4 | 56.4/63.6/59.8 | 76.2/72.2/74.1 | 79.7/79.1/79.4 |
| **FA** | 502 | 78.0/81.2/79.6 | 82.0/79.6/80.8 | 85.0/75.9/80.2 | 79.0/82.3/80.6 | 76.0/75.2/75.6 |
| **FN** | 210 | 57.1/54.5/55.8 | 54.8/59.0/56.8 | 40.5/58.6/47.9 | 57.1/55.8/56.4 | 45.2/45.2/45.2 |

Table A4. Per-Class Performance Metrics of the LSCN Network across 5-Fold Cross-Validation

| **Class** | **Samples** | **Fold 1** | **Fold 2** | **Fold 3** | **Fold 4** | **Fold 5** |
|---|---|---|---|---|---|---|
| | | **PR/SE/F1(%)** | **PR/SE/F1(%)** | **PR/SE/F1(%)** | **PR/SE/F1(%)** | **PR/SE/F1(%)** |
| **AA** | 738 | 57.3/67.5/62.0 | 53.5/63.1/58.0 | 45.1/65.4/53.4 | 53.3/55.4/54.3 | 55.1/67.5/60.7 |
| **AN** | 1852 | 79.9/77.4/78.6 | 85.1/77.7/81.2 | 82.7/75.2/78.8 | 73.4/77.1/75.2 | 82.4/77.8/80.0 |
| **BA** | 386 | 79.4/78.6/79.0 | 83.5/81.8/82.6 | 83.3/79.2/81.2 | 91.7/72.1/80.8 | 81.8/79.7/80.7 |
| **BN** | 104 | 30.8/38.1/34.1 | 42.3/52.4/46.8 | 23.1/31.6/26.8 | 7.7/20.0/11.3 | 26.6/31.6/28.9 |
| **CA** | 240 | 92.3/85.7/88.9 | 84.6/100.0/91.7 | 69.2/100.0/81.8 | 50.0/60.0/54.5 | 80.0/88.9/84.2 |
| **CN** | 87 | 78.3/88.7/83.2 | 86.7/92.9/89.7 | 90.0/88.5/89.2 | 78.3/87.0/82.4 | 83.3/83.3/83.3 |
| **DA** | 51 | 30.8/57.1/40.0 | 38.5/62.5/47.6 | 38.5/62.5/47.6 | 16.7/50.0/25.0 | 40.0/66.7/50.0 |
| **DN** | 51 | 63.6/73.7/68.3 | 45.5/62.5/52.6 | 68.2/78.9/73.2 | 71.4/65.2/68.2 | 58.8/76.9/66.7 |
| **EA** | 8129 | 99.3/97.8/98.5 | 99.1/98.0/98.5 | 99.3/98.4/98.8 | 98.9/97.7/98.3 | 99.4/97.9/98.6 |
| **EN** | 665 | 72.5/72.0/72.2 | 72.3/83.3/77.4 | 70.5/70.9/70.7 | 69.9/65.9/67.9 | 70.7/76.4/73.4 |
| **FA** | 502 | 78.6/75.0/76.8 | 81.7/75.7/78.6 | 84.8/79.1/81.9 | 82.4/77.4/79.8 | 82.0/79.6/80.8 |
| **FN** | 210 | 37.7/42.6/40.0 | 37.7/44.4/40.8 | 53.8/57.1/55.4 | 42.3/52.4/46.8 | 54.8/59.0/56.8 |

Note: PR = Precision, SE = Sensitivity, F1 = F1-Score. Bold values indicate best F1-score performance for each class. Sample sizes demonstrate the significant class imbalance, with EA (8129 samples) representing ~62% of the dataset while minority classes like DA and DN have only 51 samples each.